\documentclass[journal=jctcce,manuscript=article]{achemso}

\usepackage[version=3]{mhchem} 
\usepackage{amsmath}
\usepackage{caption}
\usepackage{subfig}
\usepackage{graphicx}
\usepackage{epstopdf}

\newcommand{\tn}{\textnormal}

\author{Kai Wang}
\author{Shiyang Long}
\author{Pu Tian}
\email{tianpu@jlu.edu.cn}
\phone{+86-431-85155287}
\affiliation
{College of Life Science}
\alsoaffiliation
{MOE Key Laboratory of Molecular Enzymology and Engineering\\
Jilin University \\
2699 Qianjin Street, Changchun 130012}

\title[]{Configurational space discretization and free energy calculation in complex molecular systems}

\abbreviations{}
\begin{document}

\section*{Abstract}
Trajectories provide dynamical information that is discarded in free energy calculations, for which we sought to design a scheme with the hope of saving cost for generating dynamical information. We first demonstrated that snapshots in a converged trajectory set are associated with implicit conformers that have invariant statistical weight distribution (ISWD). Based on the thought that infinite number of sets of implicit conformers with ISWD may be created through independent converged trajectory sets, we hypothesized that explicit conformers with ISWD may be constructed for complex molecular systems through systematic increase of conformer fineness, and tested the hypothesis in lipid molecule palmitoyloleoylphosphatidylcholine (POPC). Furthermore, when explicit conformers with ISWD were utilized as basic states to define conformational entropy, change of which between two given macrostates was found to be equivalent to change of free energy except a mere difference of a negative temperature factor, and change of enthalpy essentially cancels corresponding change of average intra-conformer entropy. These findings suggest that entropy enthalpy compensation is inherently a local phenomenon in configurational space. By implicitly taking advantage of entropy enthalpy compensation and forgoing all dynamical information, constructing explicit conformers with ISWD and counting thermally accessible number of which for interested end macrostates is likely to be an efficient and reliable alternative end point free energy calculation strategy.


\newpage
\section*{Introduction}
\begin{figure}[] 
\centering 
\includegraphics[width=5.0in]{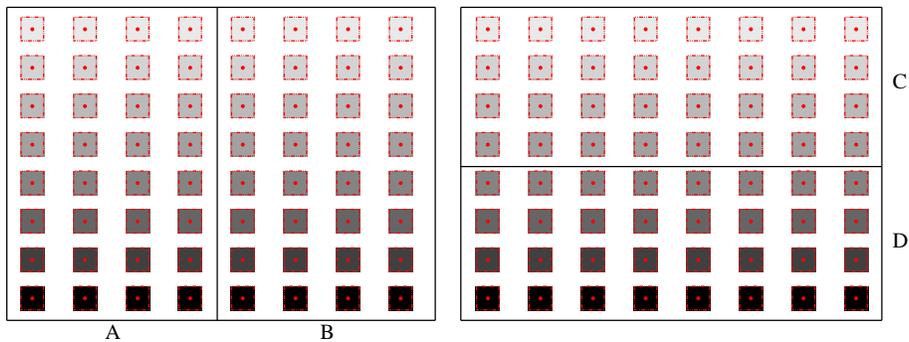}
\caption{A schematic illustration of the ISWD property in two dimension for implicit conformers associated with snapshots in a converged MD trajectory set. Red points represent snapshots, corresponding dashed squares represent associated implicit conformers with darker grayscale indicating heavier statistical weight. With shown variant statistical weight distribution of implicit conformers in the vertical direction, $\Delta F^{AB} = ln\frac{N^A_{snap}}{N^B_{snap}}$(left), while $\Delta F^{CD} \neq ln\frac{N^C_{snap}}{N^D_{snap}}$(right). As long as variation of statistical weight distribution exist, we may always find a pair of macrostates like $C$ and $D$. Therefore, robustness of the population based free energy formula (equation \ref{eq:snap}) is equivalent to the ISWD property for the corresponding set of implicit conformers.}
\label{fig:ISWD}
\end{figure}

For two arbitrary macrostates \emph{A} and \emph{B} visited in a set of converged molecular dynamics (MD) simulation trajectories, the free energy difference may be expressed as:
\begin{equation}
\Delta F^{AB} = k_BTln\frac{N^A_{snap}}{N^B_{snap}}\label{eq:snap}
\end{equation}
with $N^{A(B)}_{snap}$ being observed number of snapshots in macrostate $A(B)$, $k_B$ being Boltzmann constant and $T$ being the temperature. However, if a converged MD trajectory set was generated for the sole purpose of calculating free energy differences between interested macrostate pairs, all dynamical information contained would have been discarded. One question we sought to answer is that if there is a way to save computational cost used for generating dynamical information by designing a free energy calculation method without explicit utilization of trajectories. A rarely discussed fact is that each snapshot represents an implicit microscopic volume (termed conformer hereafter) in configurational space. More importantly, equation (\ref{eq:snap}) implies that, in a set of converged trajectories, implicit conformers associated with snapshots have \emph{invariant statistical weight distribution} (ISWD) across the whole configurational space (see Fig. \ref{fig:ISWD}). Therefore, one way to answer our original question is to accomplish the following two tasks: i) to construct a set of configurational-space-filling \footnote{Let the volume of the whole configurational space of a $N$-atom molecular system being $V_{3N}$, for a set of $M$ conformers each has a non-overlapping volume $v_i (i = 1, 2, \cdot, M)$, if $\sum^M_{i=1}v_i = V_{3N}$, then this set of conformers are configurational-space-filling.} explicit conformers, with thermally accessible ones among which have the property of ISWD ( or a sufficiently good approximation of it ), and ii) to design an efficient method to count such conformers that are thermally accessible in given macrostates. To be concise, we use ``explicit conformers with ISWD (ECISWD)'' to represent ``configurational-space-filling explicit conformers, with thermally accessible ones among which have the property of ISWD ( or a sufficiently good approximation of it )'' hereafter.  For two arbitrary macrostates $A$ and $B$ that have $N_{conf}^A$ and $N_{conf}^B$ (Note that both are functions of potential energy) thermally accessible conformers, denoting corresponding average statistical weight of conformers as $w^A$ and $w^B$, the change of free energy between these two macrostates may be written as:
\begin{align}
\Delta F^{AB} & = k_BTln\frac{N_{conf}^Aw^A}{N_{conf}^Bw^B} \nonumber \\
                       & = k_BTln\frac{N_{conf}^A}{N_{conf}^B} + k_BTln\frac{w^A}{w^B}
\end{align}
For ECISWD, $w^A \approx w^B$, therefore:
\begin{equation}
\Delta F^{AB} \approx k_BTln\frac{N_{conf}^A}{N_{conf}^B}\label{eq:conf}
\end{equation}
It was demonstrated that sequential Monte Carlo (SMC) in combination with importance sampling\cite{zhang2003,zhang2006} may rapidly count the number of explicit conformers that are thermally accessible. Therefore, the hinging issue is to construct a set of ECISWD. We set to address this issue and accompanying implications in this study.

\section*{Hypothesis on ECISWD}
Conformers associated with MD snapshots are implicit with no information available for their shapes or sizes, we consequently may not directly learn from MD trajectories. One principal consideration for defining ECISWD is sufficient fineness since statistical weight of complex molecular systems are in general exponentially different for different macrostates\cite{Skilling2006}, very coarse conformers are associated with the possibility that the heavist conformer in the statistically most dominant macrostate weighs more than the total of all other macrotates, hence rendering ISWD impossible. Better uniformity is another factor to consider for the same reason. It is noted that ISWD holds for each set of implicit conformers associated with snapshots of corresponding independent and converged MD trajectory set. Therefore, infinite number of ways exist for constructing sets of implicit conformers with ISWD for a given complex molecular systems. Based on this thought, we hypothesized that any set of sufficiently fine and uniform conformers should approximately have the property of ISWD, and we may consequently define ECISWD through systematically increasing their fineness according to our convenience. 

\begin{figure}[] 
\centering 
\includegraphics[width=5.0in]{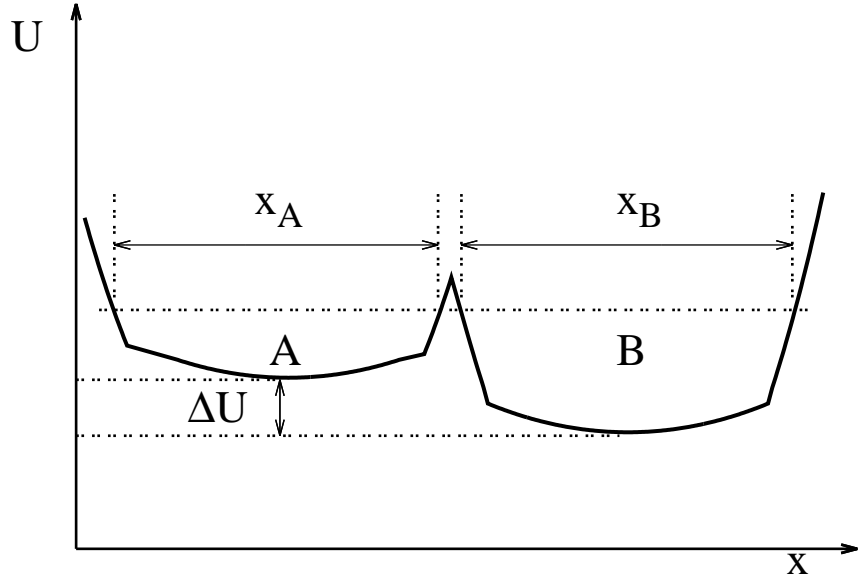}
\caption{A simple double well potential with equal width (i.e. $x_A = x_B$). $U$ is the potential energy and $\Delta U$ is the potential energy difference between the macrostates $A$ and $B$.}
\label{fig:doublewell}
\end{figure}

This hypothesis is immediately disproved by a simple double well system shown in Fig. \ref{fig:doublewell}. With increasingly different $\Delta U$ between two wells $A$ and $B$, regardless of the fineness for any uniformly defined conformers, the statistical weight distribution of which in two macrostates will be increasingly different. The only way to achieve sufficiently good approximate ISWD is to construct conformers that were properly weighted by $U$, the potential energy surface that we do not know \emph{a priori} in a real complex molecular system. Nonetheless, complex molecular systems are very different from a double well system. As shown in Fig. \ref{fig:doublewell}, if we divide macrostates $A$ and $B$ into $N^A$ and $N^B$ (e.g. $N^A = N^B = 1000,000$ ) conformers,  $U$ is consistently higher in $A$ than in $B$ in terms of conformer average, and within each conformer $U$ is essentially a constant. Such situation is unlikely, if ever possible, to occur in a complex molecular system. With large number of degrees of freedom (DOFs), tight packing and steep van der waals repulsive core of constituting atoms, potential energy may vary significantly within a microscopic volume of configurational space. Therefore, we think that competitions among large number of DOFs may render construction of ECISWD an achievable task, and the above mentioned hypothesis may well be valid for complex molecular systems. 

Sufficiently well-converged MD trajectory sets of specific molecular systems provide ideal test grounds for ISWD property of given explicit conformers based on the following two arguments. Firstly, trajectory sets are generated by known force fields, and therefore no convolution of force fields inaccuracy and experimental error exists as in the case of comparing computational results with experimental ones; Secondly, we may arbitrarily partition configurational space visited in a trajectory set, and a hypothesis tested for arbitrarily given partitions should remain true for the whole configurational space. This is an important logic since traversing configurational space for complex molecular systems is practically impossible. The symbolic equivalence between equation (\ref{eq:snap}) and equation (\ref{eq:conf}) suggests that for a set of ECISWD, if we assign each snapshot in a trajectory set to a corresponding conformer and utilizing equations (\ref{eq:snap}) and (\ref{eq:conf}) respectively to calculate free energy changes for arbitrarily selected pairs of macrostates, differences in results caused by different conformer definitions ( between a given explicit conformer set and the implicit one associated with snapshots ) should decrease with increasing size of trajectory set and essentially disappear for a fully converged trajectory set, the reason is that free energy difference between two arbitrarily given macrostates does not depend on the way it is calculated. Conversely, if statistical weight distribution of a set of explicit conformers is widely different in different part of the configurational space, the corresponding differences in results would increase with increasing size of trajectory set and saturate for a fully converged trajectory set since the largest possible error is limited by the number of available snapshots in any trajectory sets that are not fully converged. Both complete disappearance of differences resulted from equations (\ref{eq:snap}) and (\ref{eq:conf}) for the case of ECISWD, and full saturation of differences resulted from these two equations for the case of explicit conformers without ISWD will be extremely difficult to observe for complex molecular systems duo excessive amount of data needed. Nonetheless, the trend should be equivalently informative as long as the largest trajectory set is sufficiently well-converged.

We chose lipid POPC to carry out such tests based on the fact that large MD trajectory sets are available for this molecule. Specifically, we firstly extracted MD trajectories of POPC from trajectories of M2 muscarinic acetylcholine receptor study\cite{Dror2013}. Three increasingly larger trajectory sets, TSA1, TSA2 and TSA3 were constructed with smaller trajectory sets being subsets of larger ones. Secondly, we defined four different sets of conformers, which were denoted as CONF1 through CONF4 (see Fig. \ref{fig:POPC}) respectively, with CONF1 being the finest and CONF4 being the coarsest. Thirdly, we used backbone dihedrals as order parameters to construct macrostates through projection operations. Finally, number of conformers ($N_{conf}$) were calculated for each macrostate of the given combination of trajectory set and definition of conformers (see \emph{Methods} for details). 

\begin{figure}
\centering 
\subfloat[]{\includegraphics[width=1.6in]{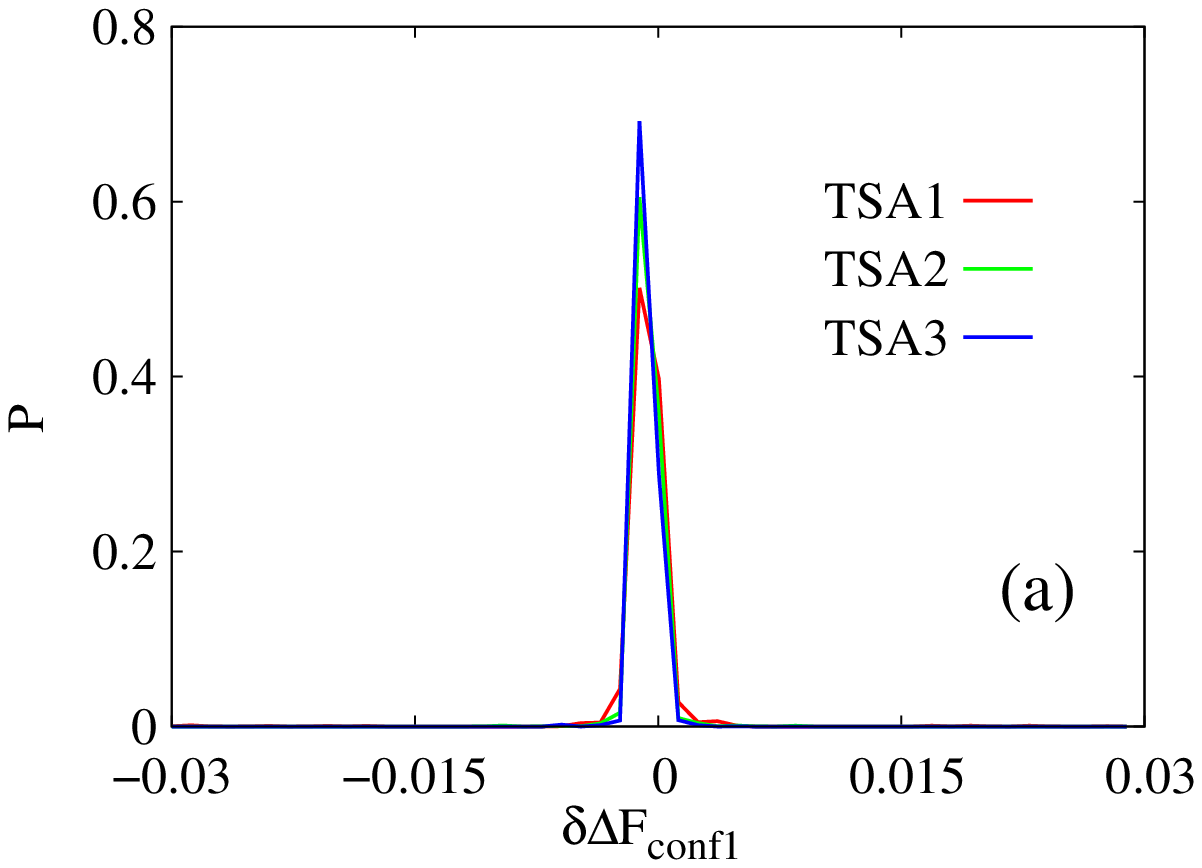}}
\subfloat[]{\includegraphics[width=1.6in]{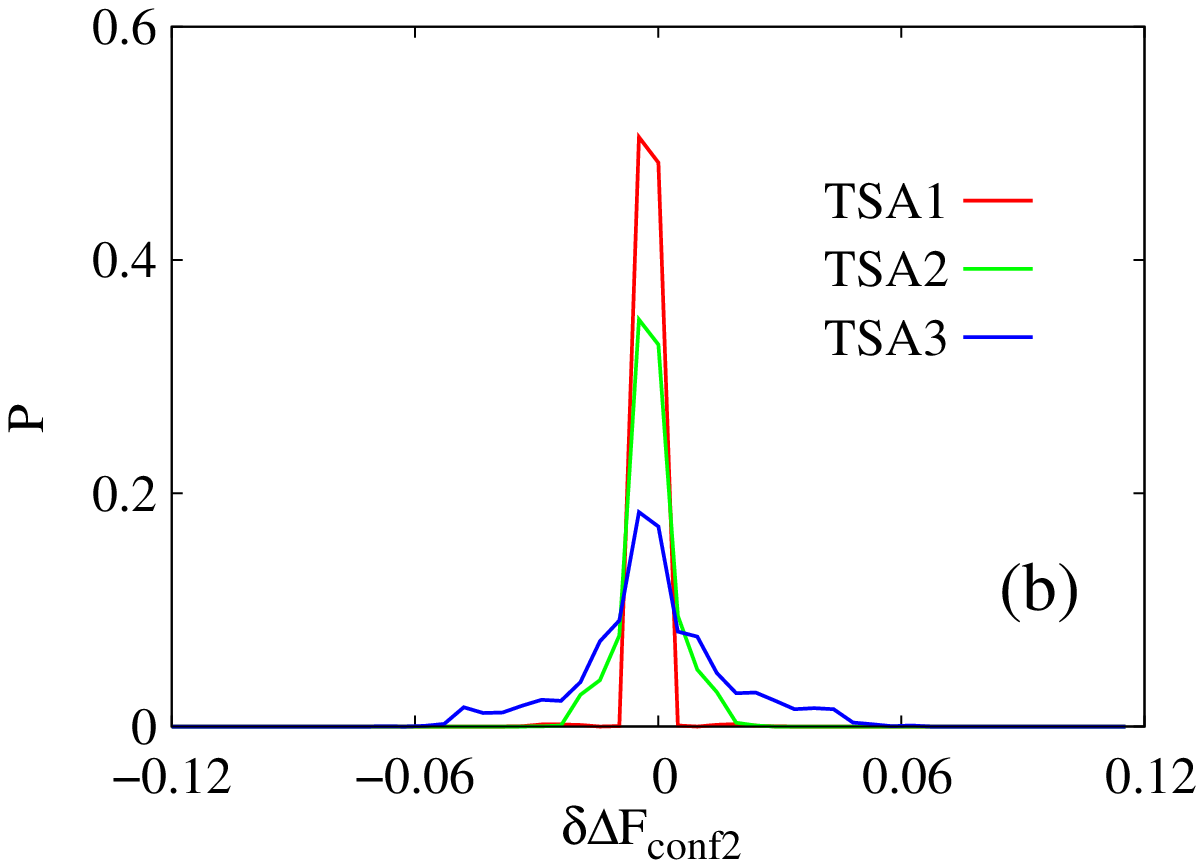}}
\subfloat[]{\includegraphics[width=1.6in]{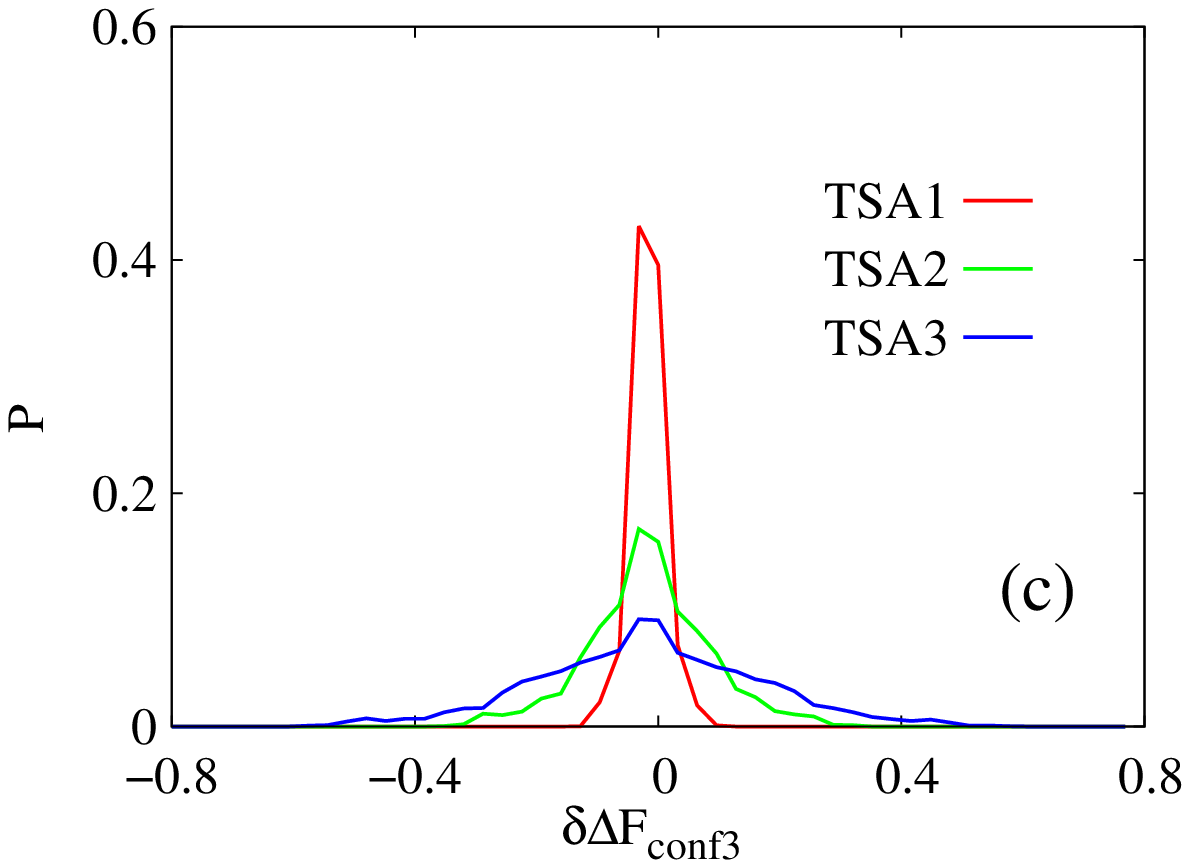}}
\subfloat[]{\includegraphics[width=1.6in]{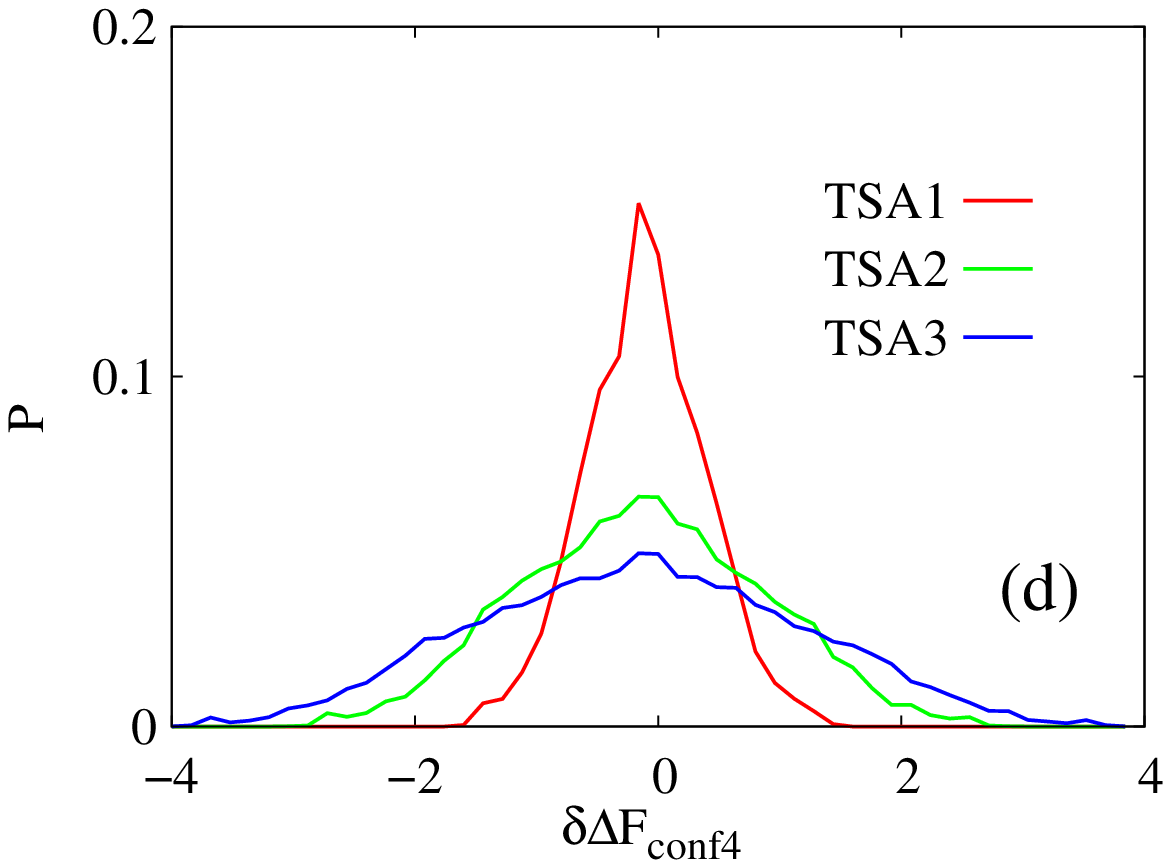}}\\
\subfloat[]{\includegraphics[width=1.6in]{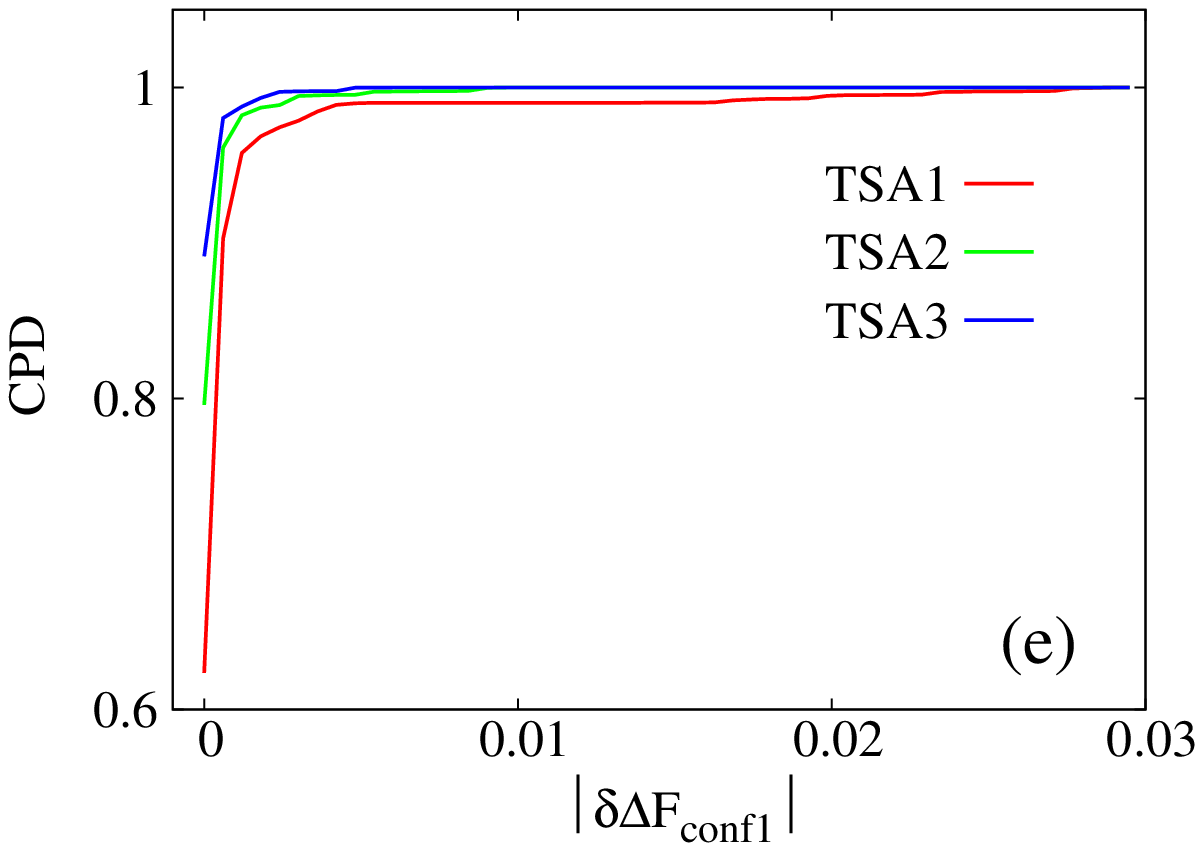}}
\subfloat[]{\includegraphics[width=1.6in]{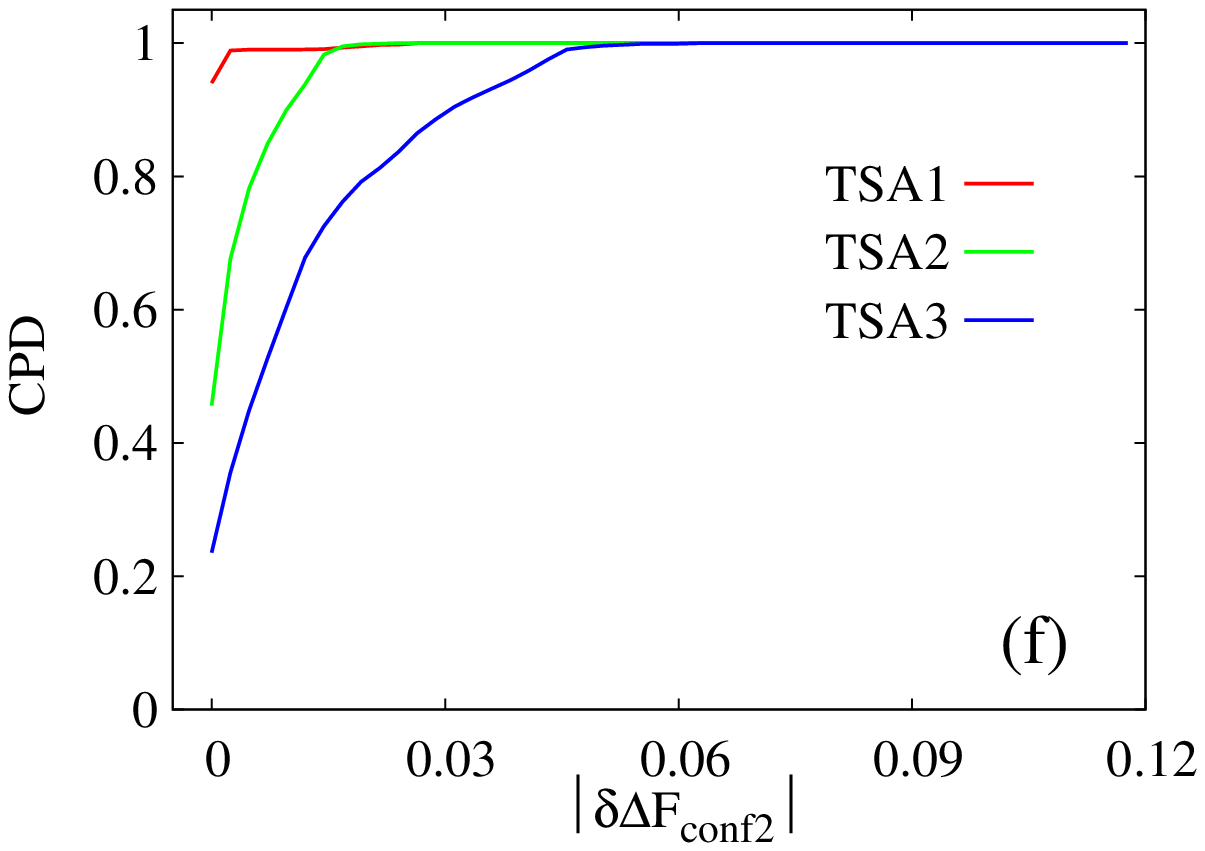}}
\subfloat[]{\includegraphics[width=1.6in]{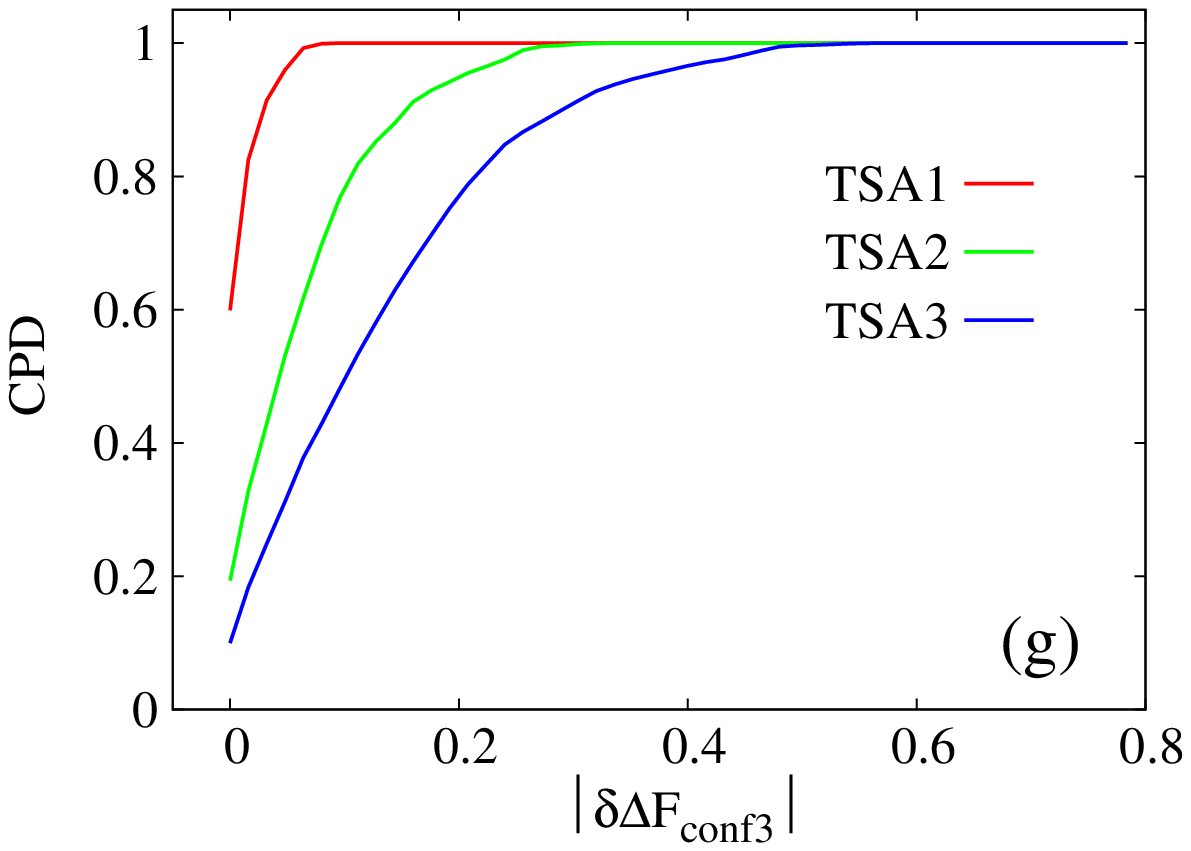}}
\subfloat[]{\includegraphics[width=1.6in]{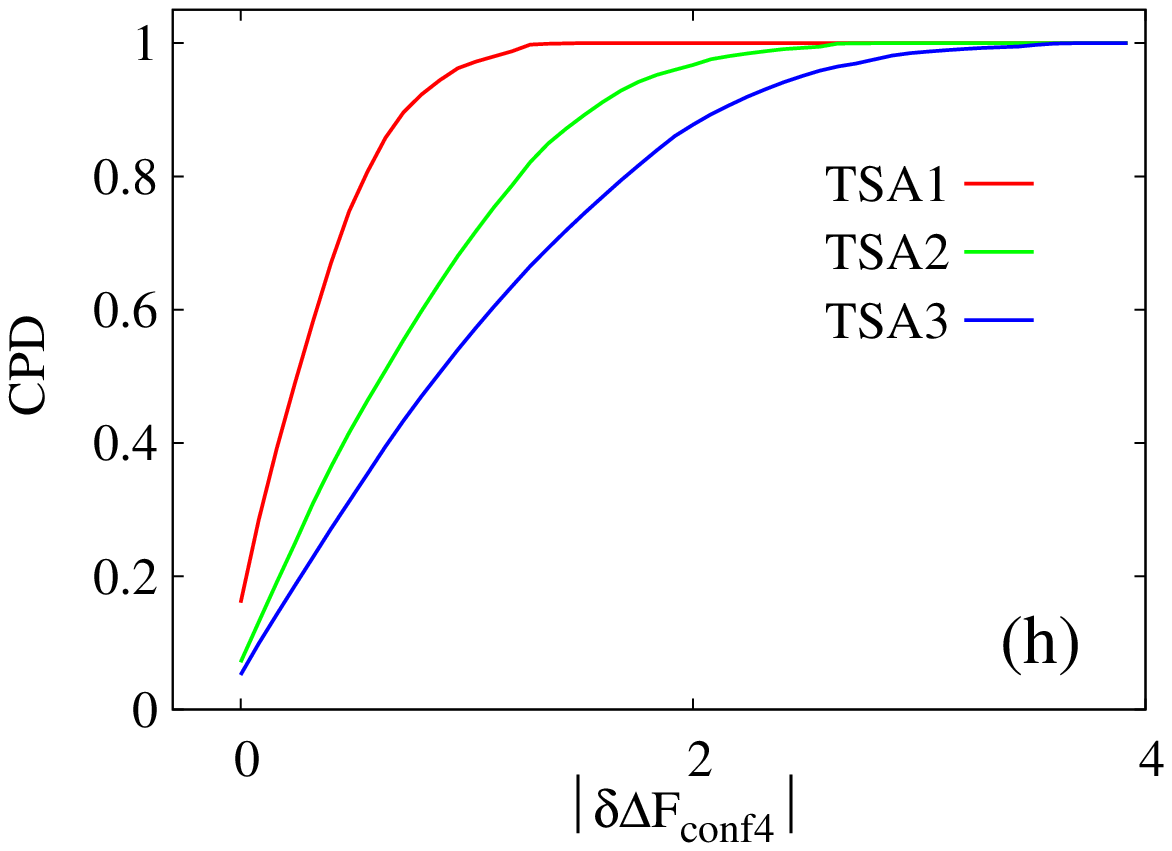}}\\
\caption{Distributions of $\delta\Delta F$ (a - d) and CPD  of its absolute values (e - h) for POPC with four sets of explicit conformers (CONF1 through CONF4, which are indicated in the horizontal label as subscripts, e.g. $\delta\Delta F_{conf1}$ in (a) and $|\delta\Delta F_{conf1}|$ in (e)). Different trajectory sets are represented by different line colors. The unit of the horizontal axis is in $k_BT$.} 
\label{fig:CPD}
\end{figure}

\begin{figure}[] 
\centering 
\subfloat[]{\includegraphics[width=1.6in]{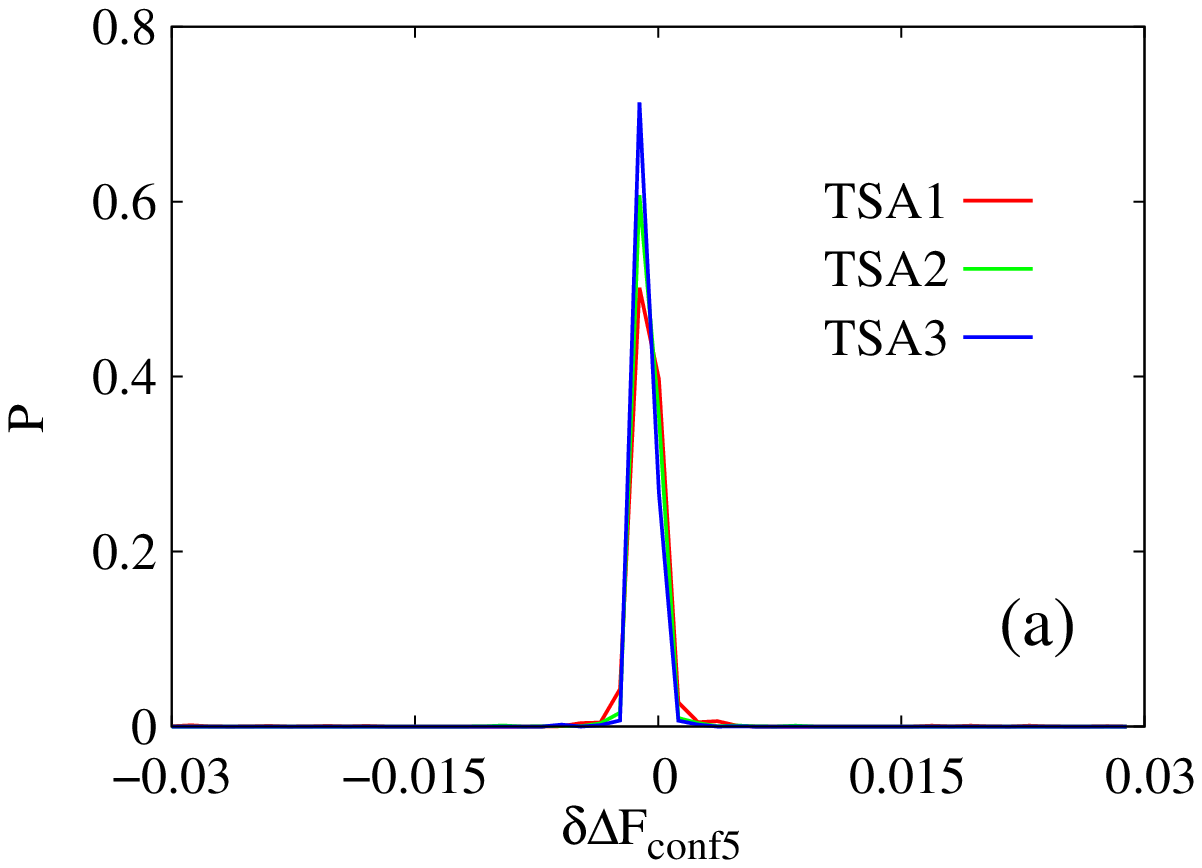}}
\subfloat[]{\includegraphics[width=1.6in]{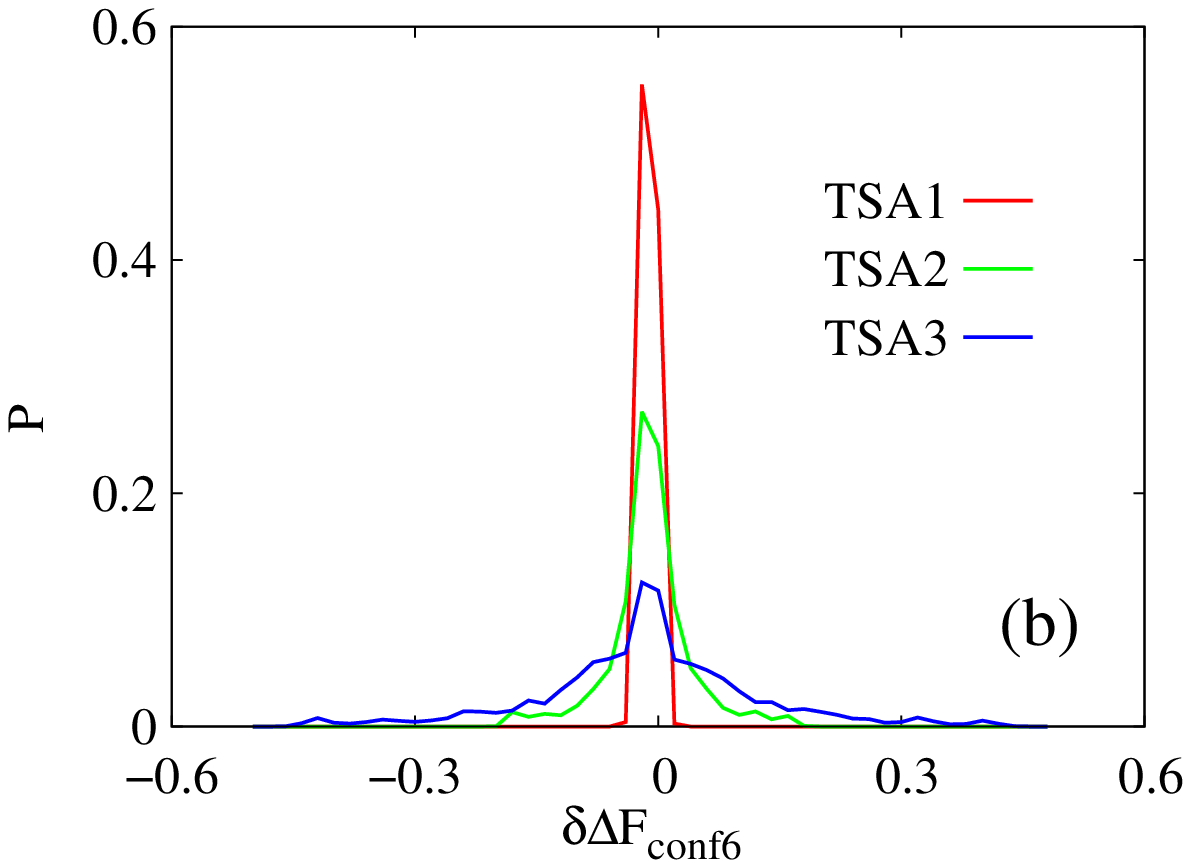}}
\subfloat[]{\includegraphics[width=1.6in]{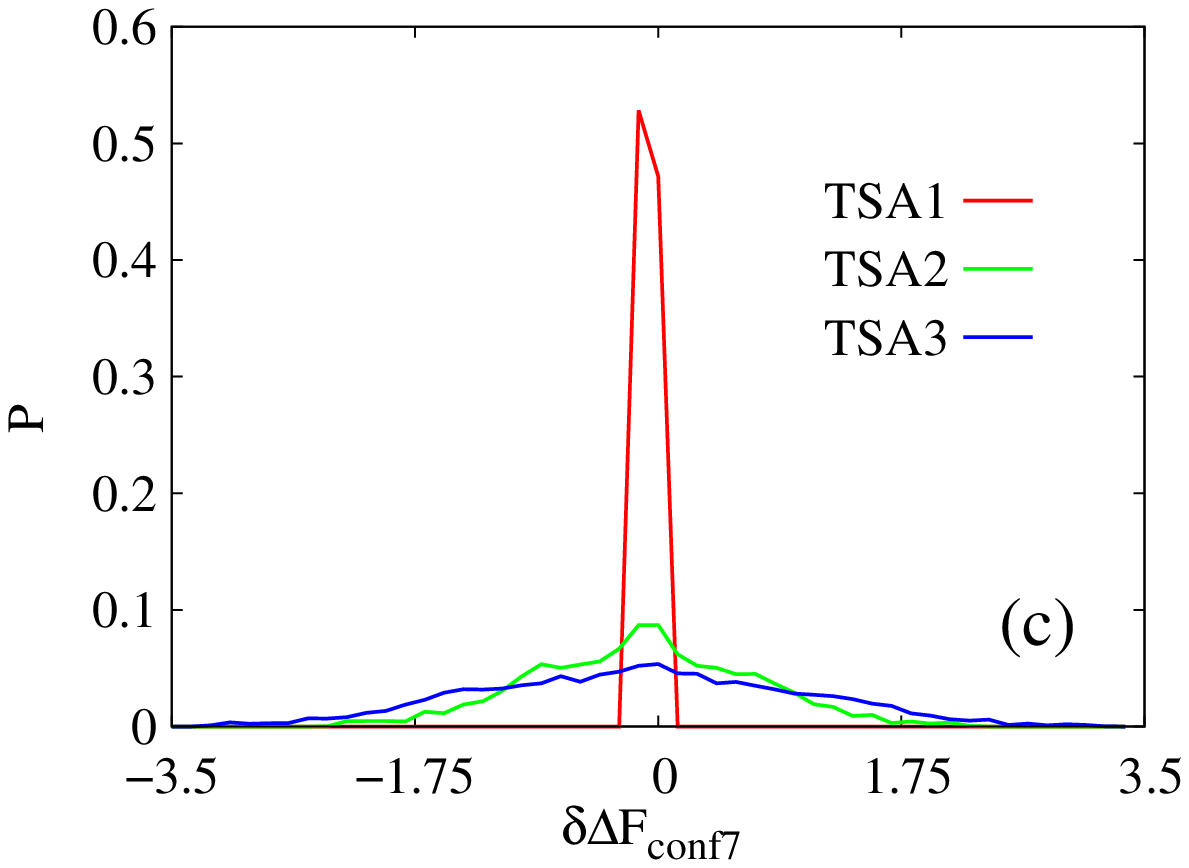}}
\subfloat[]{\includegraphics[width=1.6in]{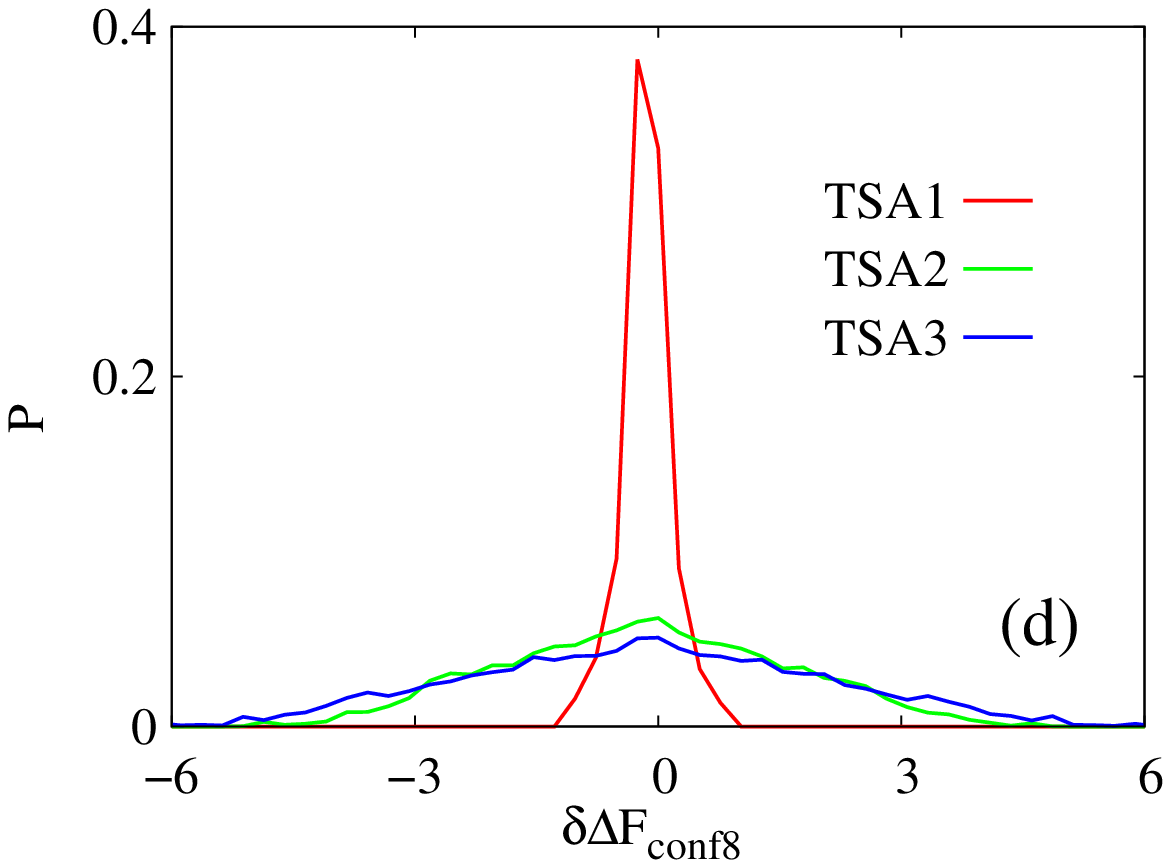}}\\
\subfloat[]{\includegraphics[width=1.6in]{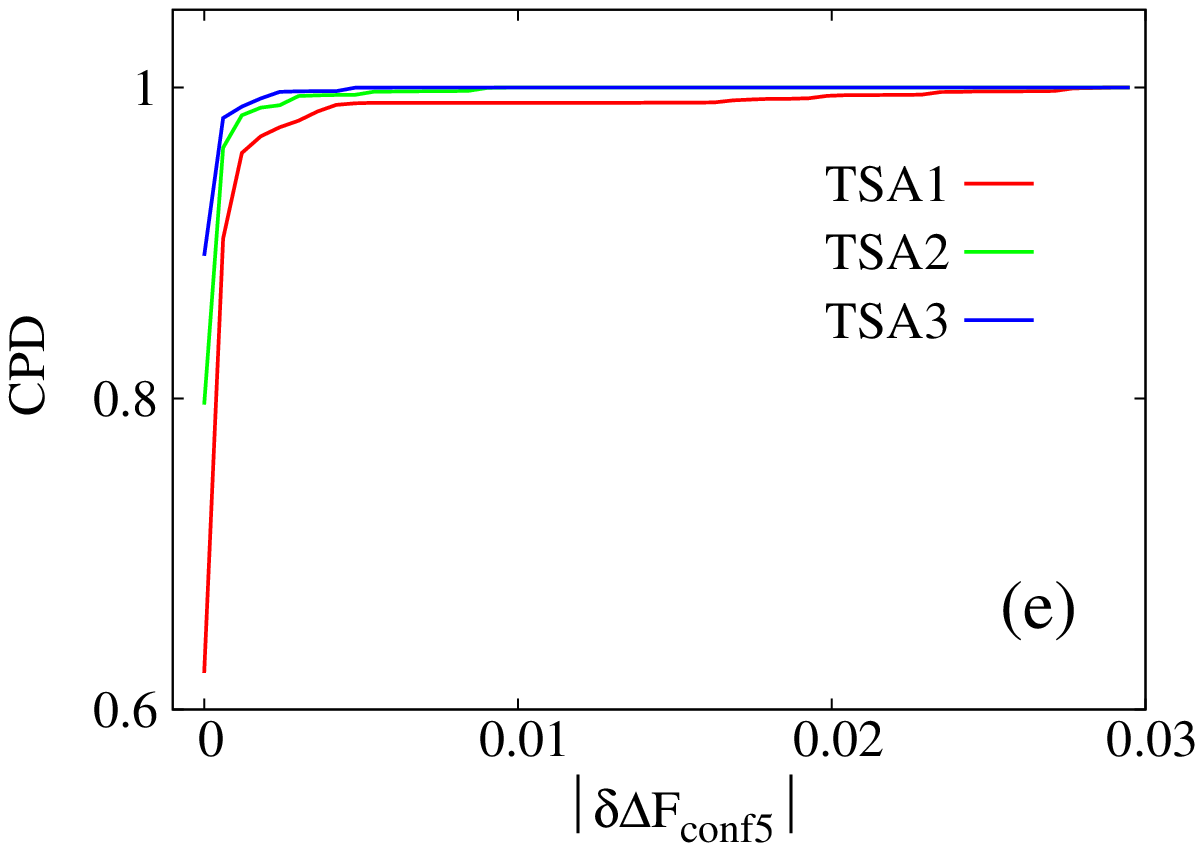}}
\subfloat[]{\includegraphics[width=1.6in]{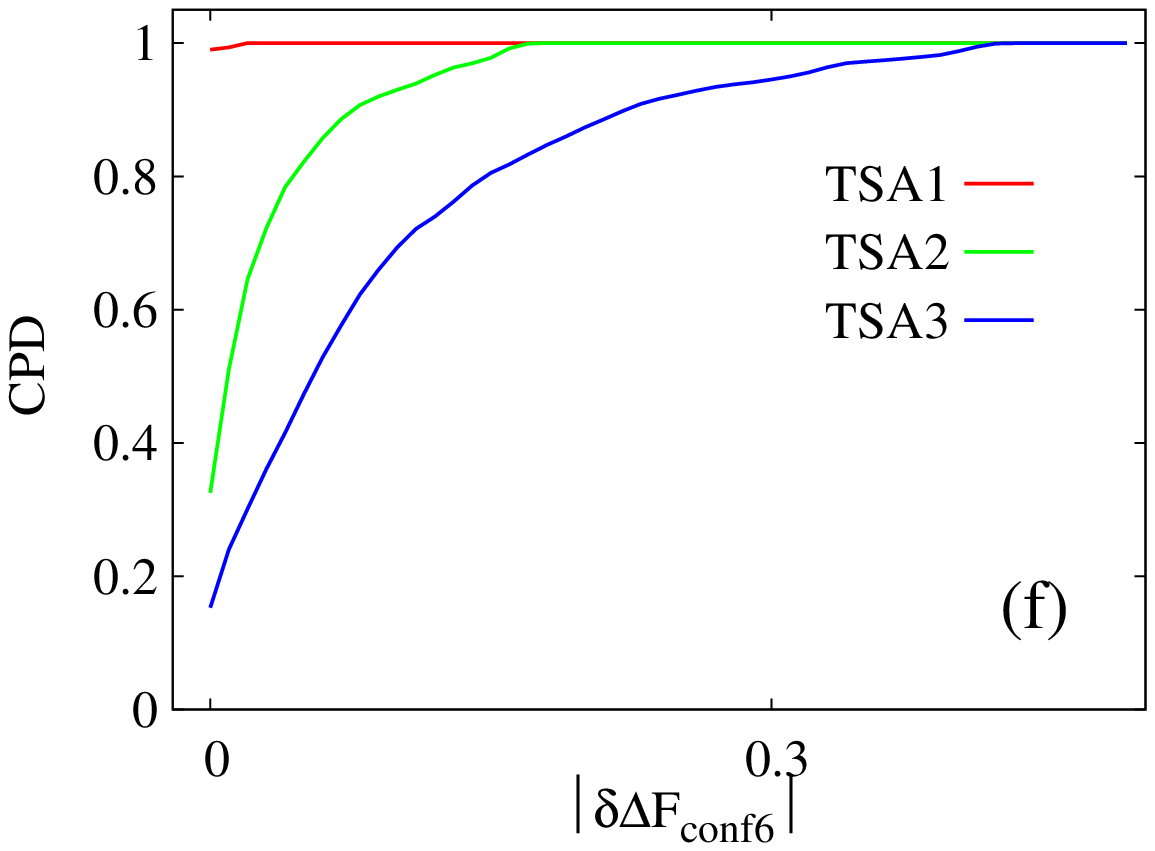}}
\subfloat[]{\includegraphics[width=1.6in]{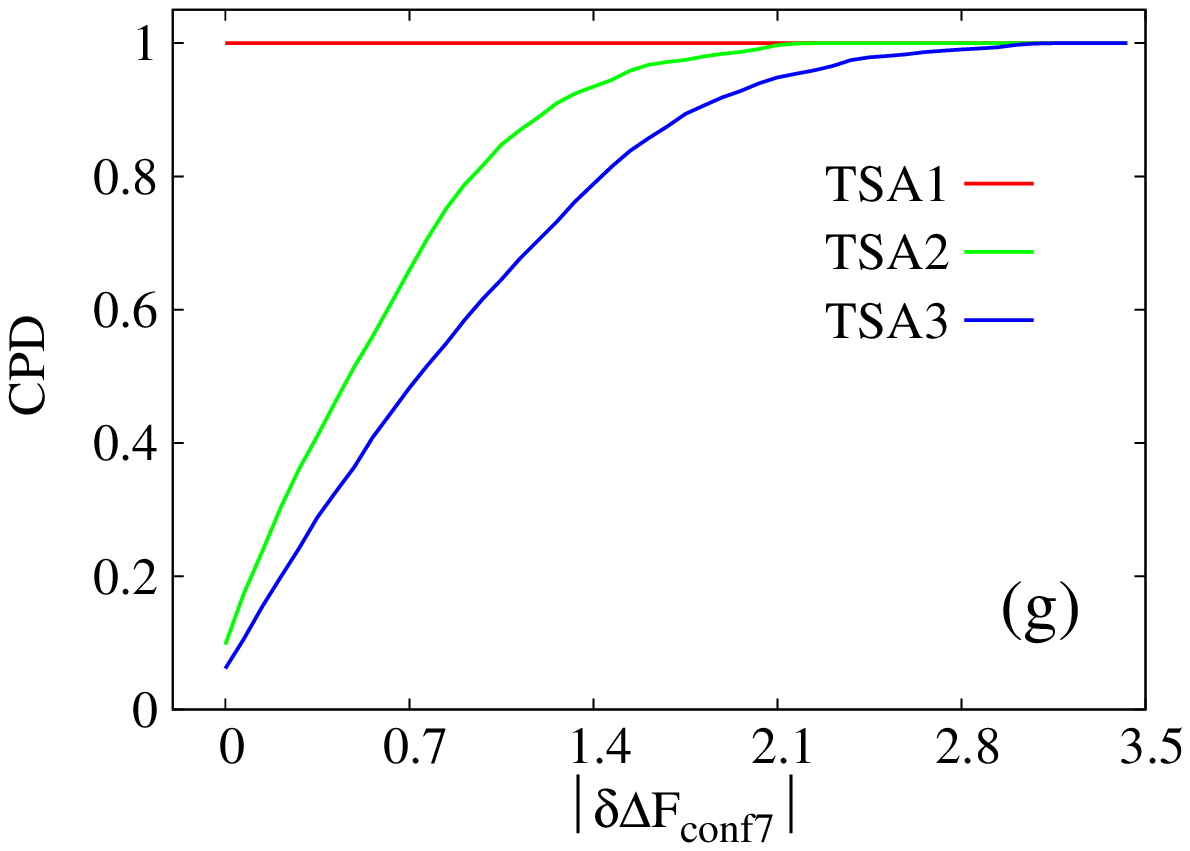}}
\subfloat[]{\includegraphics[width=1.6in]{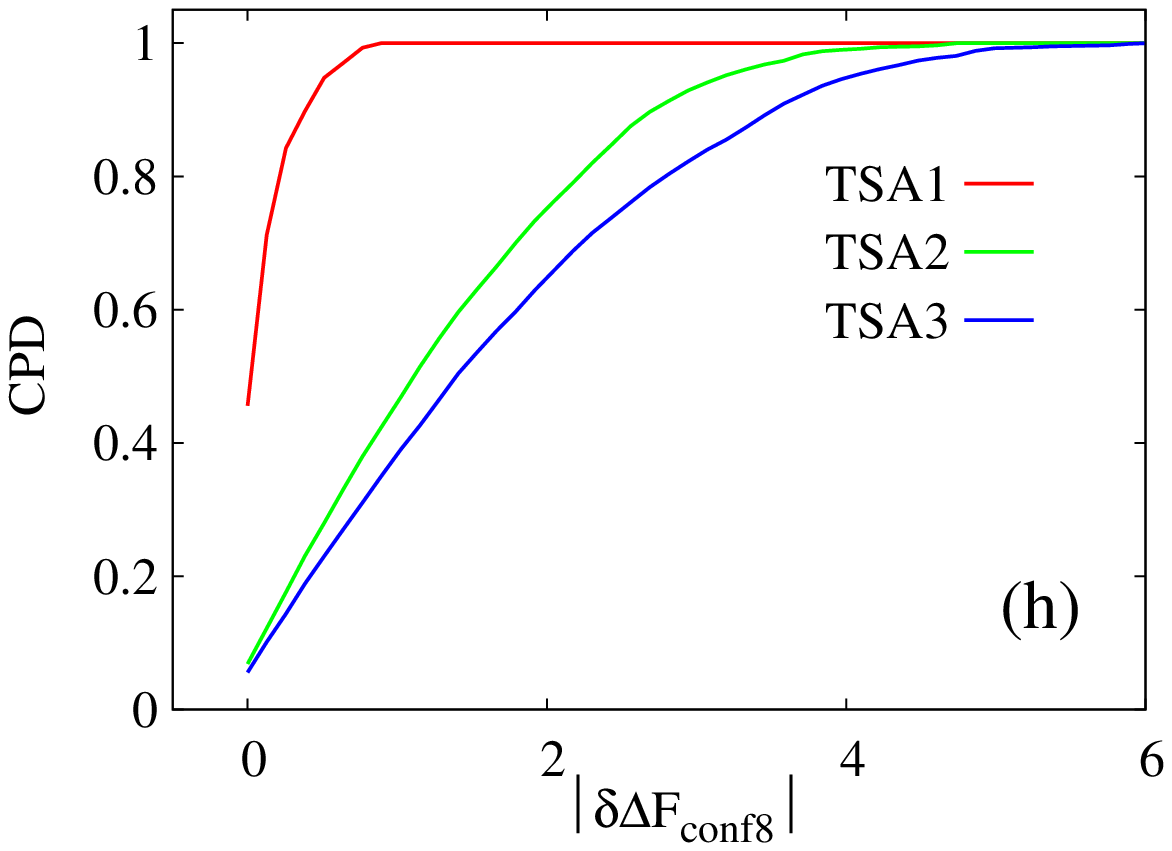}}\\
\caption{Distributions of $\delta\Delta F$ (a - d) and CPD  of its absolute values (e - h) for POPC with conformer sets CONF5 through CONF8, which are defined similarly to CONF1 through CONF4 except that torsional states boundaries are $60^{\circ}$, $180^{\circ}$ and $300^{\circ}$. Different trajectory sets are represented by different line colors. The unit of the horizontal axis is in $k_BT$.} 
\label{fig:A5678}
\end{figure}

With the above given definitions of conformers, macrostates and trajectory sets, we calculated $\Delta F$ for all pairs of macrostates on each combination of conformer definition and trajectory set according to equation (\ref{eq:snap}) (denoted as $\Delta F_{snapshot}$) and equation (\ref{eq:conf}) (denoted as $\Delta F_{conformer}$) respectively, and their differences were denoted as $\delta\Delta F = \Delta F_{snapshots} - \Delta F_{conformer}$, which essentially measures differences between our constructed set of explicit conformers and implicit conformers associated with snapshots. Distributions of $\delta\Delta F$ and cumulative probability density (CPD) of its absolute values for the four sets of explicit conformers (CONF1 through CONF4) are shown in Fig. \ref{fig:CPD}. Firstly, for CONF2 through CONF4 (Fig. \ref{fig:CPD}b-d), distribution of $\delta\Delta F$ is significantly broader for larger trajectory set. Secondly, it is noted that the range of horizontal axis is widely different for these three sets of conformers (ranging from less than 0.1$k_BT$ to a few $k_BT$).  For a given trajectory set, dramatically broader distribution of $\delta\Delta F$ is observed for coarser conformer definitions. Correspondingly, CPD plots of $\delta\Delta F$ (Fig. \ref{fig:CPD}f-h) exhibit the extent of errors more directly. These observations match our expectation for coarse conformers that do not have sufficiently good approximation of ISWD. Finally and most importantly, for CONF1 (Fig. \ref{fig:CPD}a), distribution of $\delta\Delta F$ is narrower for larger trajectory set, and is significantly narrower than that of all other conformers (Fig. \ref{fig:CPD}b-d), the CPD plot (Fig. \ref{fig:CPD}e) shows the differences among trajectory sets more clearly. Therefore, conformers in set CONF1 match our expectation for ECISWD. The observation of the behavior for CONF1 through CONF4 suggest that, as hypothesized, we may define a set of ECISWD through systematic increase of conformer fineness. Regarding the uniformity of conformers, we equally partitioned each torsional DOF into three torsional states since we have no better information \emph{a priori} to divide otherwise. To test further the hypothesis that any sufficiently fine conformers should have similarly good approximation of ISWD, we defined a few more different set of conformers with similar fineness to CONF1 through CONF4 respectively, and similar observations were made (see Fig. \ref{fig:A5678}). On different trajectory sets of POPC with similar size to TSA1 through TSA3, similar observations were made (see Fig. \ref{}fig:TSB). It is noted that regardless of conformer definition and trajectory set size, distributions of $\delta\Delta F$ is approximately symmetric with the mode at zero (Fig. \ref{fig:CPD}a-d, Fig. S1 a-d and Fig. S2 a-d), this is inevitable since selection of start and end macrostate is arbitrary and consistent in calculating both $\Delta F_{snapshot}$ and $\Delta F_{conformer}$.   

\begin{figure}
\centering
\subfloat[]{\includegraphics[width=1.6in]{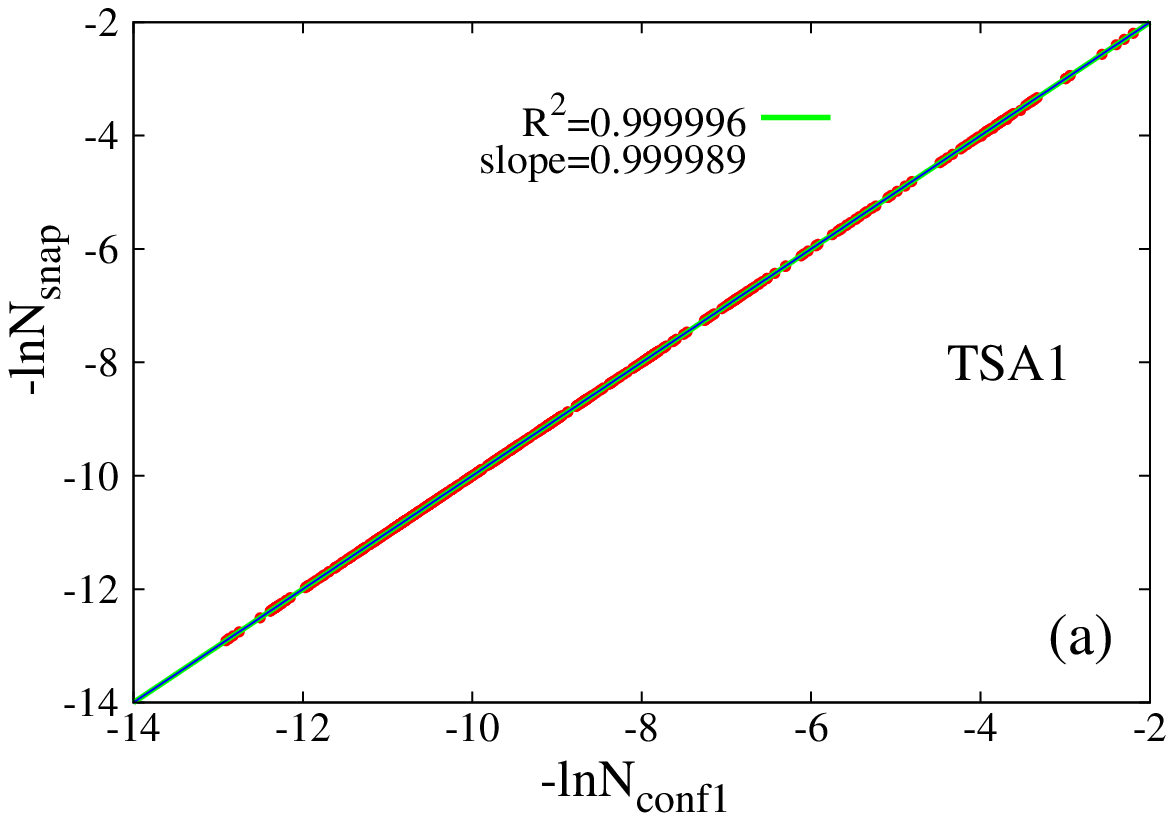}}
\subfloat[]{\includegraphics[width=1.6in]{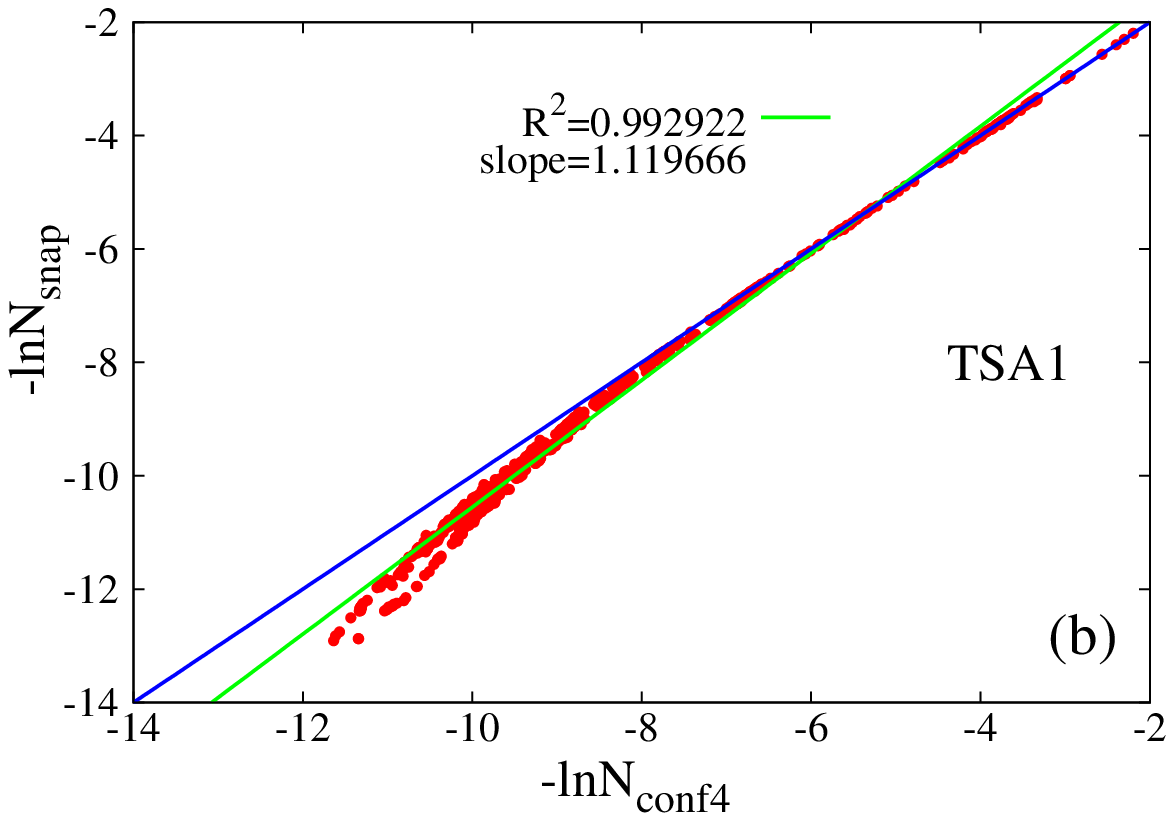}}\\
\subfloat[]{\includegraphics[width=1.6in]{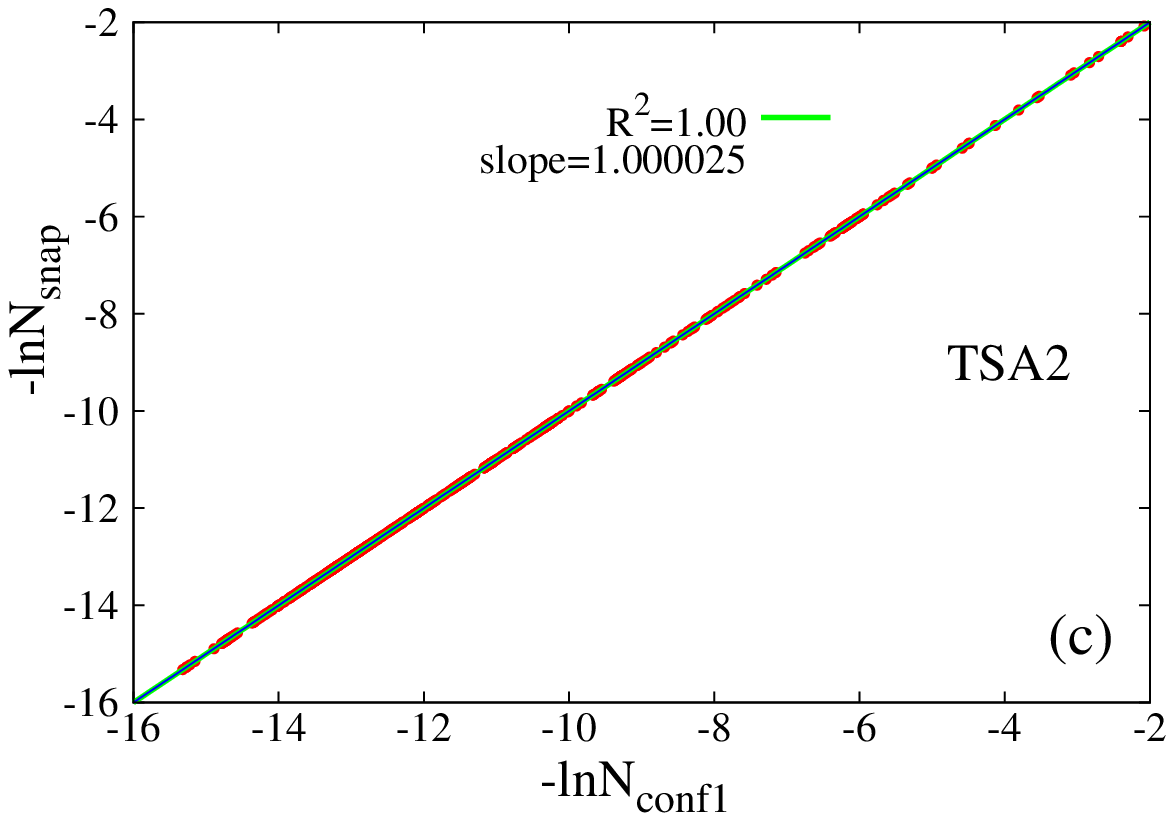}}
\subfloat[]{\includegraphics[width=1.6in]{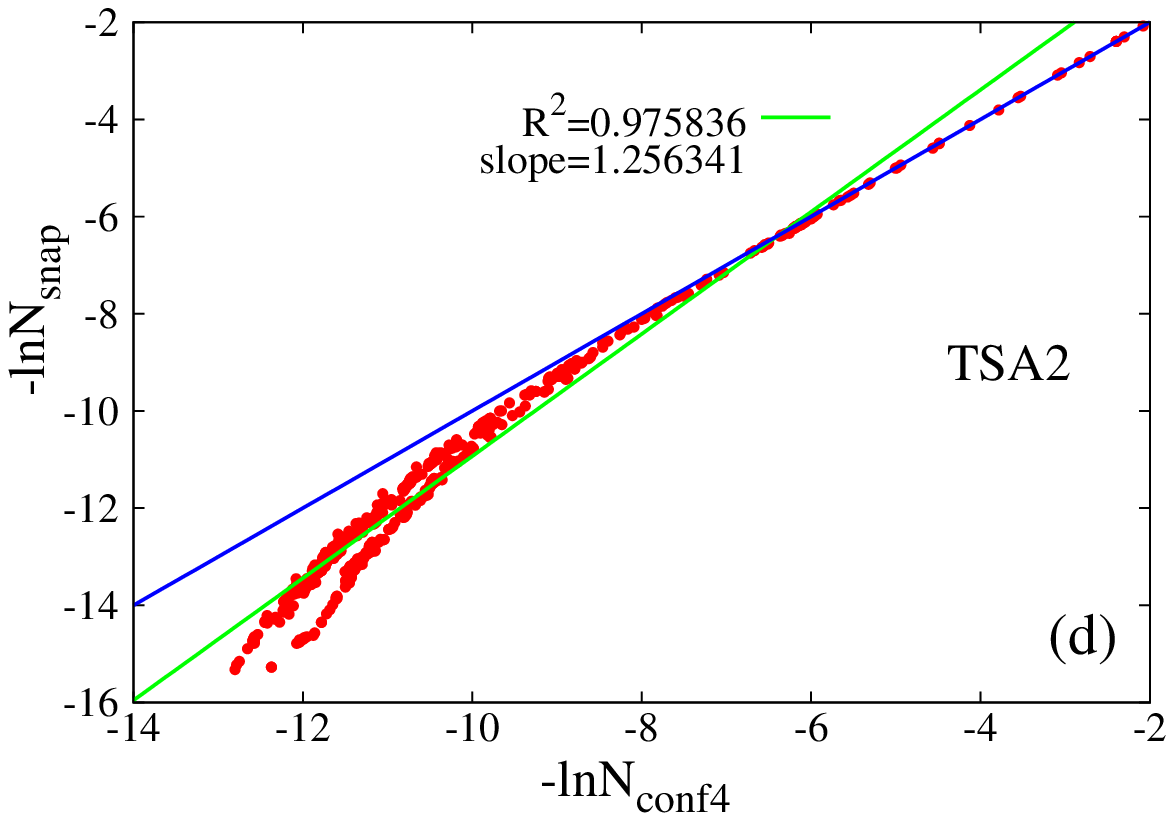}}\\
\subfloat[]{\includegraphics[width=1.6in]{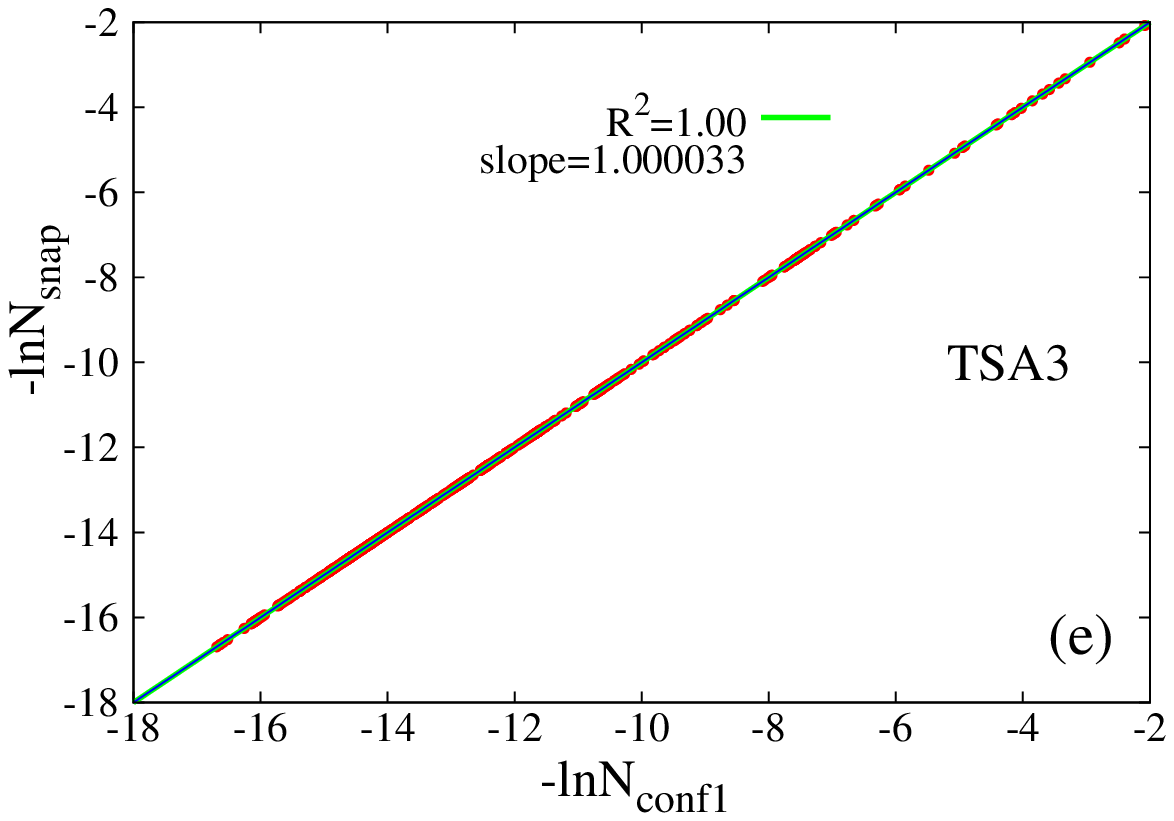}}
\subfloat[]{\includegraphics[width=1.6in]{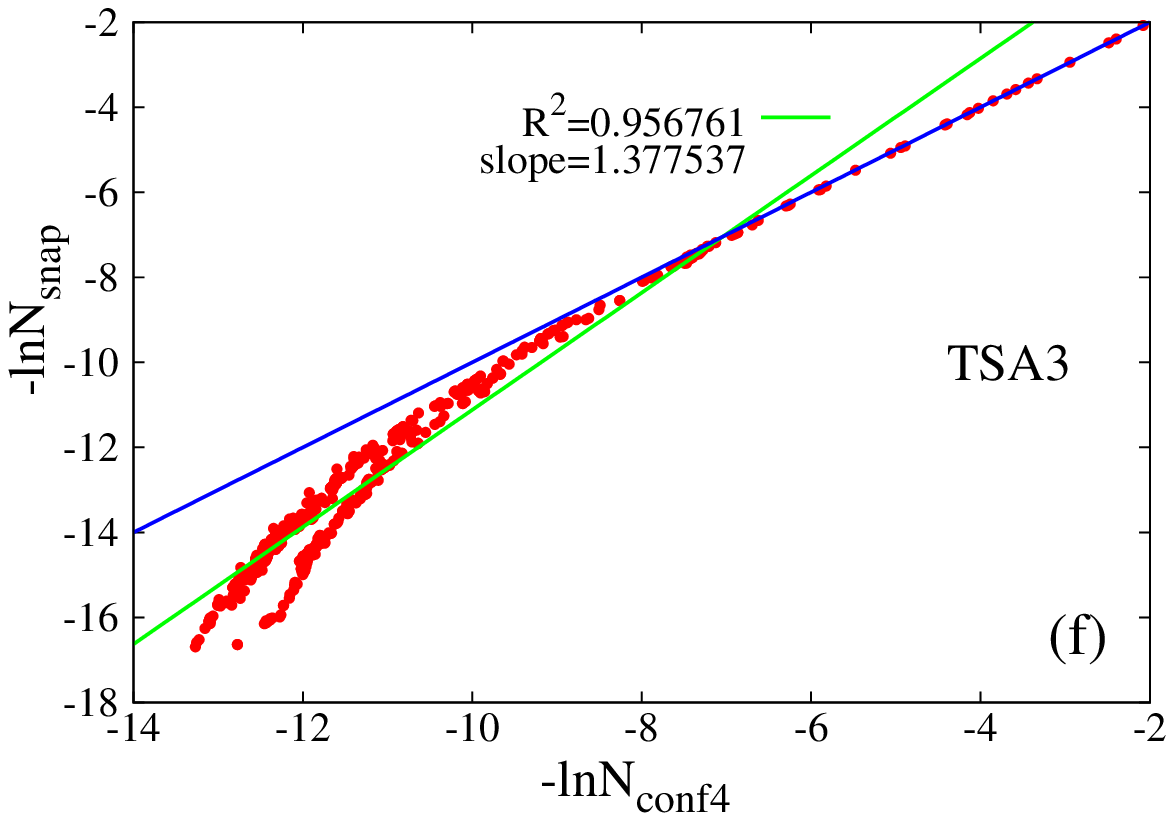}}\\
\caption{The $-lnN_{snap}\quad vs. -lnN_{conf}$ plots for CONF1 (left) and CONF4 (right) on the three trajectory sets. Blue lines represent situations where equation (\ref{eq:conf}) holds sufficiently well. Each red dot represents a macrostate; green lines are the best linear fits for the observed data with $R^2$ being the squared linear correlation coefficients.}
\label{fig:ff}
\end{figure}

For coarser explicit conformers without ISWD, deviations from ISWD are expected to occur in the heaviest macrostates, where larger probability for occurrence of excessively heavy conformers would cause uneven distribution of statistical weight. Again, such deviations are expected to be larger for larger trajectory sets (and eventually saturate for a fully converged trajectory set). To this end, we plotted $-lnN_{snap}$ vs $-lnN_{conf}$ for all constructed macrostates in Fig. \ref{fig:ff} for CONF1 and CONF4. Indeed, deviations occur for the heaviest macrostates and are larger for larger trajectory set for CONF4 (Fig. \ref{fig:ff}b,d,f). Perfect scaling was observed for CONF1 (Fig. \ref{fig:ff}a,c,e) as expected. 

\section*{Conformational entropy based on ECISWD}
Typical molecular systems in chemical, materials and biological studies, when treated quantum mechanically, present intractable complexity. Classical (continuous) representation of atomic DOFs, however, presents an awkward situation for the definition of microstates and entropy\cite{Wehrl1978}. Correspondingly, density of states of classical systems may be determined only up to a multiplicative factor\cite{Series}. The term ``conformational entropy'', despite its widespread usage, has no well established definition available for major complex biomolecular systems. Explicit conformers with ISWD, despite its system dependence and the fact that infinite number of specific definitions exist for each given complex molecular systems, may be utilized as basic states for defining conformational entropy in an abstract and general sense for any complex molecular systems, and we explore this idea and its implications in this section.     

\begin{figure}[] 
\centering 
\subfloat[]{\includegraphics[width=1.6in]{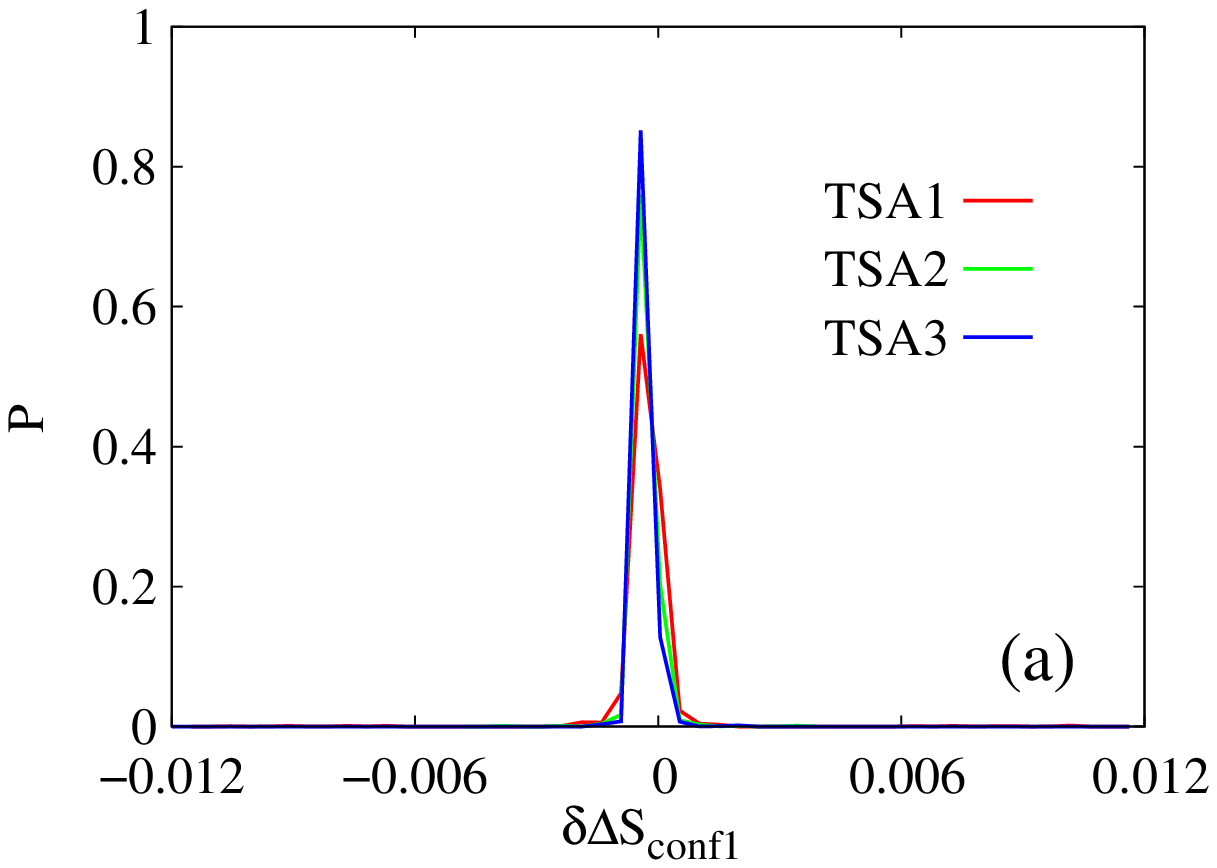}}
\subfloat[]{\includegraphics[width=1.6in]{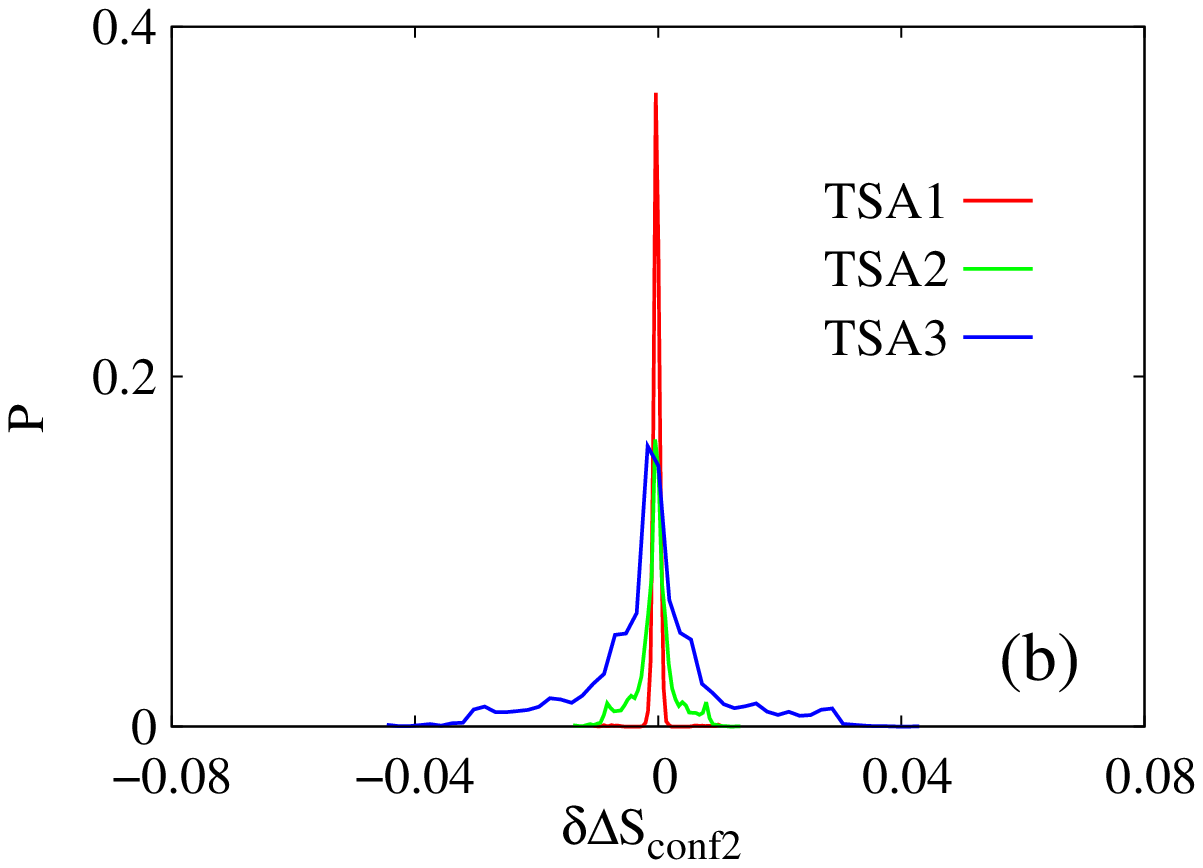}}
\subfloat[]{\includegraphics[width=1.6in]{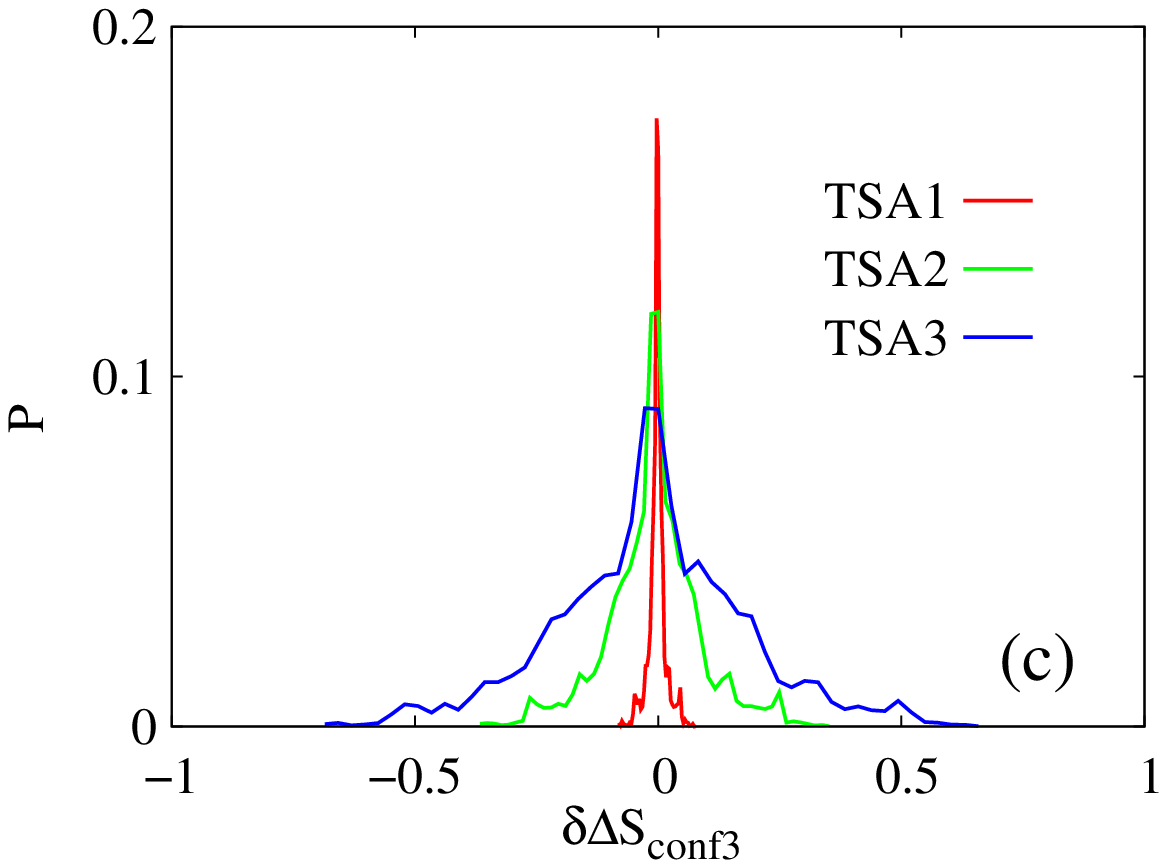}}
\subfloat[]{\includegraphics[width=1.6in]{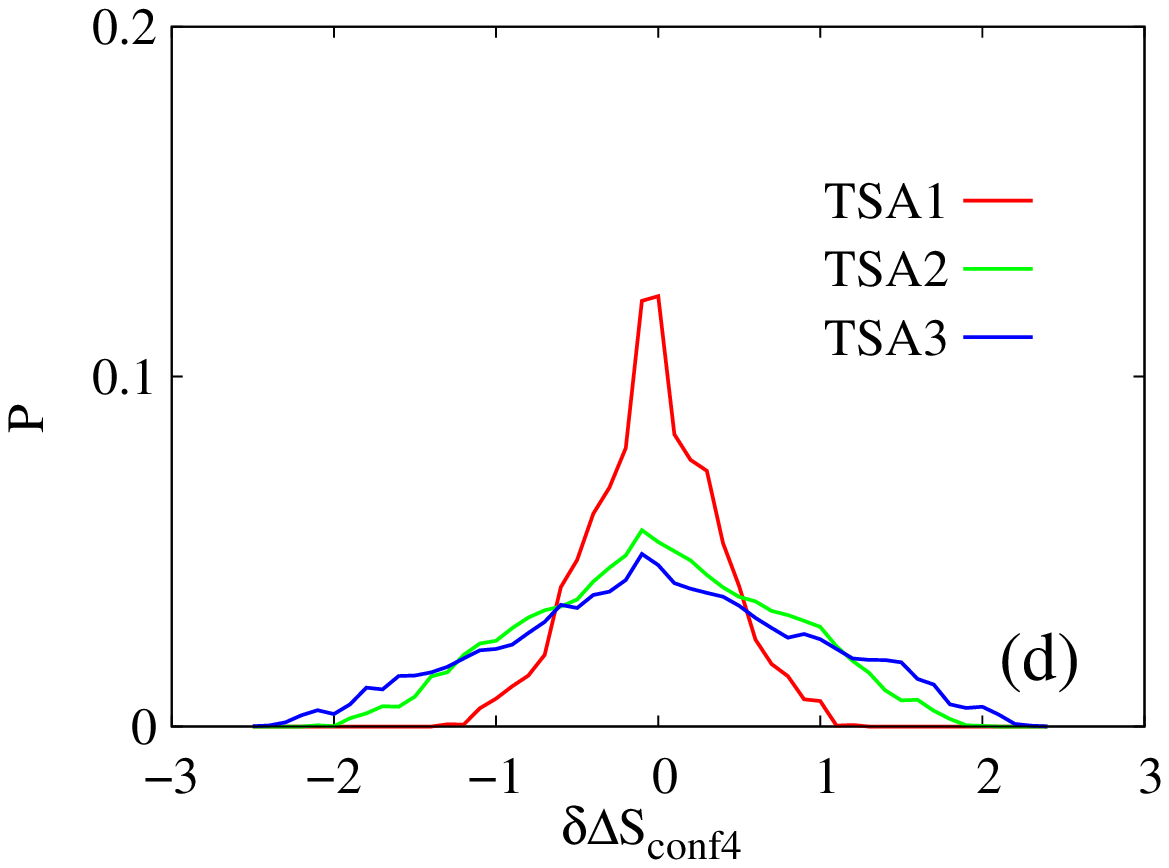}}\\
\subfloat[]{\includegraphics[width=1.6in]{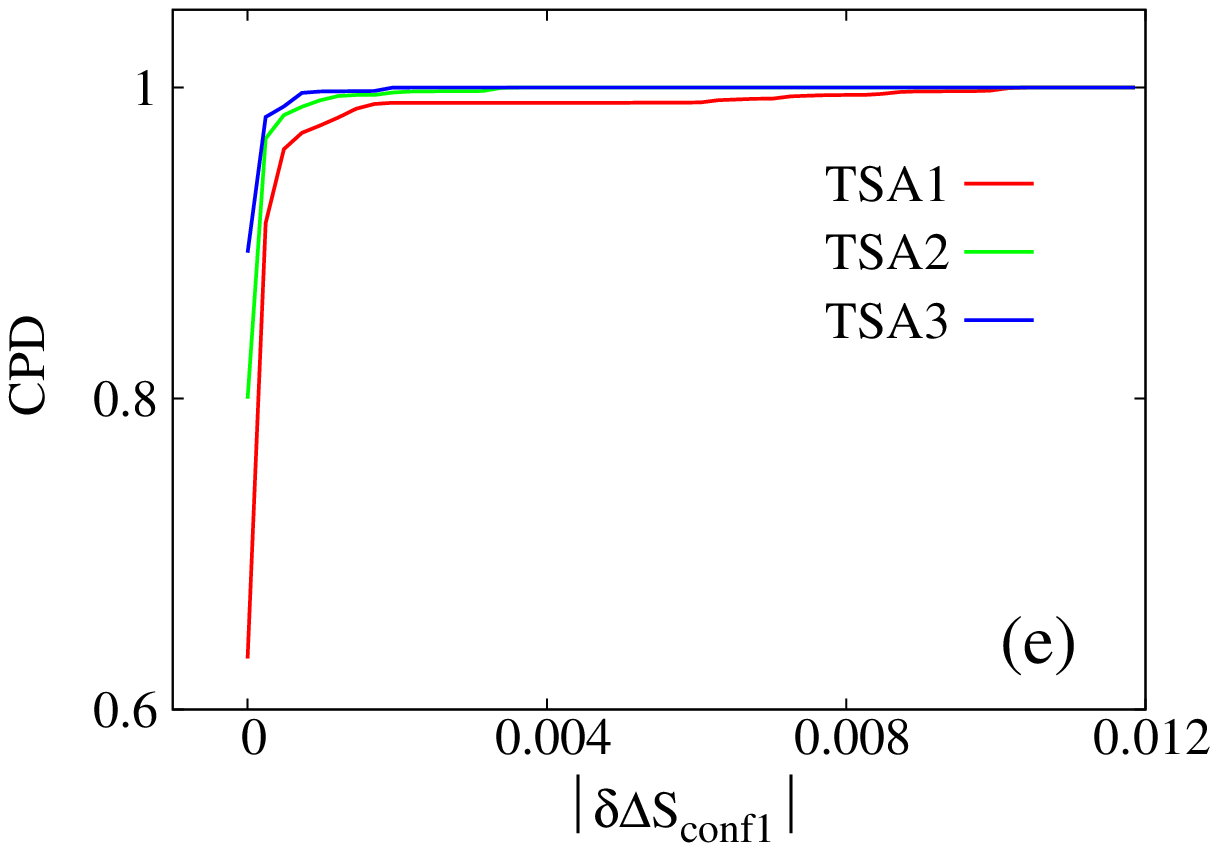}}
\subfloat[]{\includegraphics[width=1.6in]{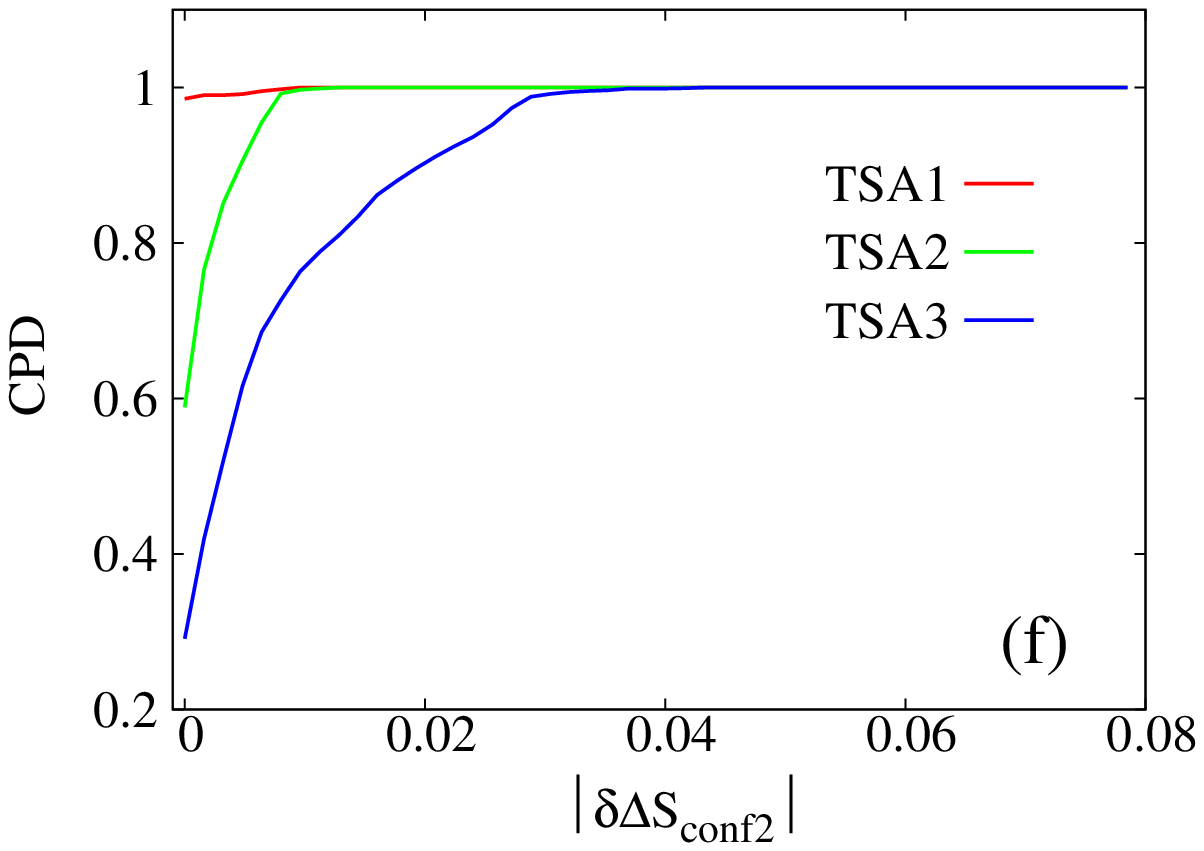}}
\subfloat[]{\includegraphics[width=1.6in]{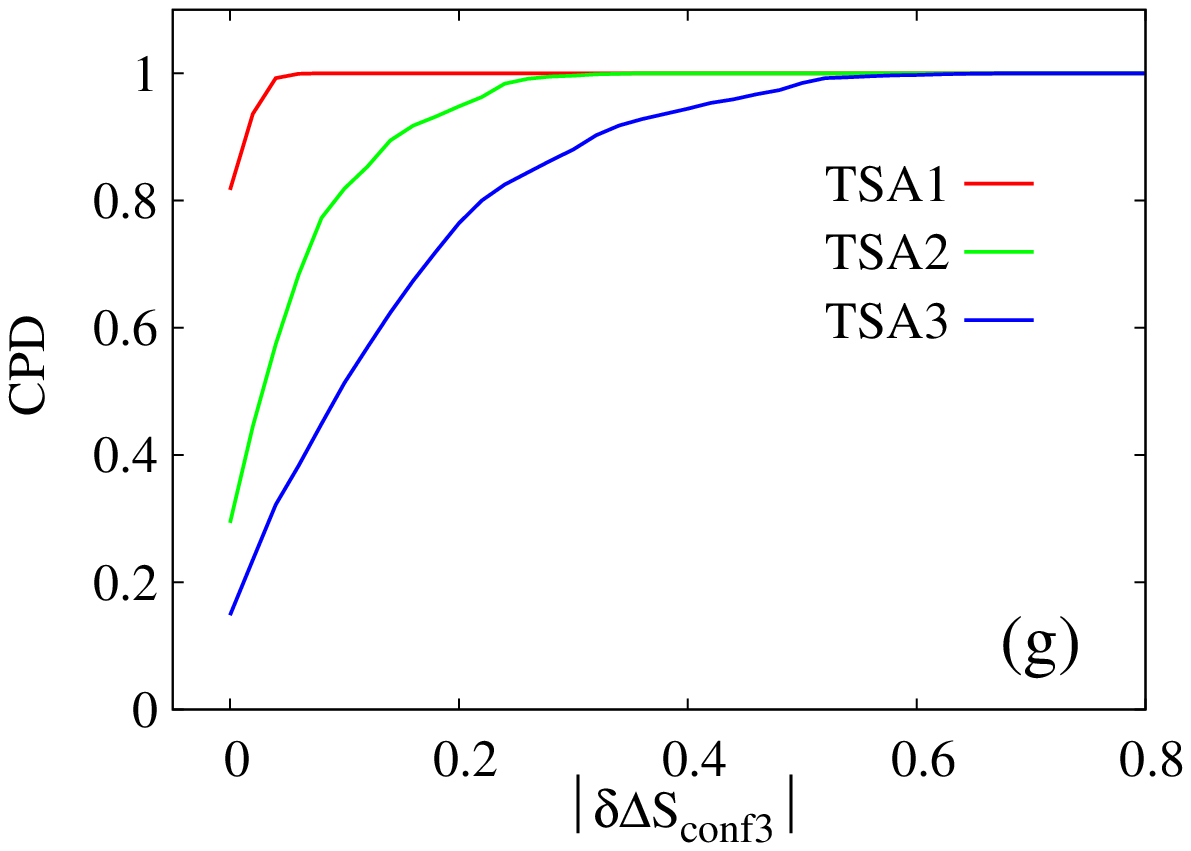}}
\subfloat[]{\includegraphics[width=1.6in]{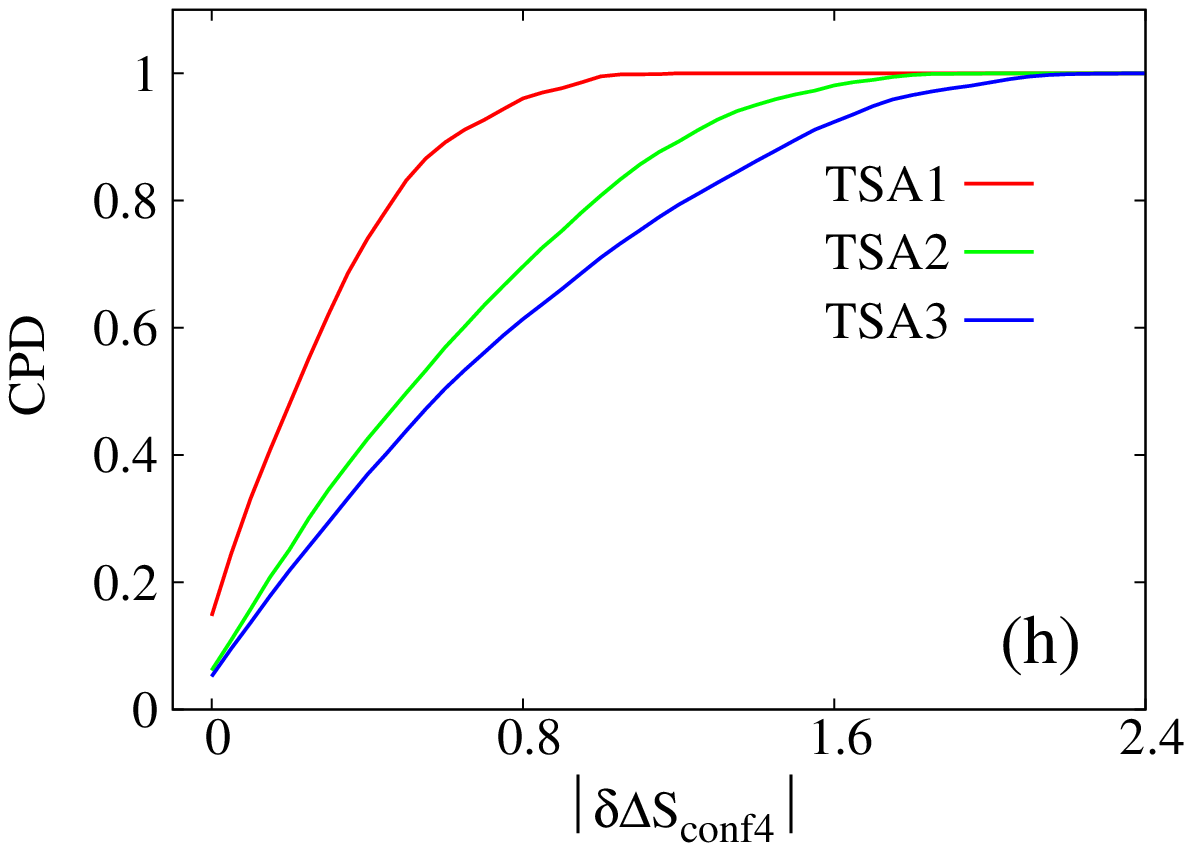}}\\
\caption{Distributions of $\delta\Delta S_{conf}$ (a - d) and CPD  of its absolute values (e - h) for POPC with four sets of explicit conformers (CONF1 through CONF4, which are indicated in the horizontal label as subscripts, e.g. $\delta\Delta S_{conf1}$ in (a) and $|\delta\Delta S_{conf1}|$ in (e)). Different trajectory sets are represented by different line colors. The unit of the horizontal axis is in $k_B$.} 
\label{fig:SCPD}
\end{figure}

It is well established in the informational theory field\cite{Shannon1948} that for a given static distribution with well-defined basic states, entropy may be constructed by arbitrary division of the whole system into $M$ subparts. 
\begin{align}
S &= -\sum^{i=N}_{i=1} P_ilnP_i = -\sum^{j=M}_{j=1}P_jlnP_j + \sum^{j=M}_{j=1} P_jS_j \\
S_j &= -\sum^{k=k_j}_{k=1}P_klnP_k   \quad (j = 1, 2, \cdot\cdot\cdot ,M) \\
N & = \sum^{j=M}_{j=1}k_j
\end{align}
with $P_i$, $P_j$ and $P_k$ being properly normalized:
\begin{equation}
\sum^{i=N}_{i=1}P_i = 1, \quad \sum^{j=M}_{j=1}P_j = 1 \quad \tn{and} \quad \sum^{k=k_j}_{k=1}P_k = 1 \quad (j = 1, 2, \cdot\cdot\cdot ,M)
\end{equation}
$S$ is the global informational entropy and $S_j$s $(j = 1, 2, \cdot\cdot\cdot ,M)$ are local informational entropies, it is noted that such division may be carried out recursively. 
We may similarly construct both local entropies of macrostates (say $A$ and $B$) and global entropy for the given molecular system based on a set of explicit conformers:
\begin{align}
S^A & = -k_B\sum_{j=1}^{N_{conf}^A}p_jlnp_j + \sum_{j=1}^{N_{conf}^A}p_jS_j^A \label{astot} \\
S^B & = -k_B\sum_{k=1}^{N_{conf}^B}q_klnq_k + \sum_{k=1}^{N_{conf}^B}q_kS_k^B \label{bstot} \\
S & = -k_B\sum_{i=1}^NP_ilnP_i + \sum_{i=1}^{N}P_iS_i \label{stot}
\end{align}
$P_i$ is the probability of the $i$th conformer in the global configurational space, $p(q)_{j(k)}$ is the probability of the $j(k)$th conformer in macrostate $A(B)$. $S_i$ is the intra-conformer entropy of the $i$th conformer in the global configurational space. $S^{A(B)}_{j(k)}$ is the intra-conformer entropy for the $j(k)$th conformer in macrostate $A(B)$. Again, $P_i$, $p_j$ and $q_k$ are properly normalized:
\begin{equation}
\sum^{i=N}_{i=1}P_i = 1, \quad \sum^{j=N_{conf}^A}_{j=1}p_j = 1 \quad \tn{and} \quad \sum^{k=N_{conf}^B}_{k=1}q_k = 1
\end{equation}
The first terms on the right hand side of equations (\ref{astot}, \ref{bstot} and \ref{stot}) describe distributions of conformer statistical weights within a macrostate or within the whole configurational space, and is referred to as ``conformational entropy'' ($S_{conf}$), the second terms are averages of the intra-conformer entropies of corresponding conformers and are denoted ${S_{int}}$. We may rewrite $S^A$ and $S^B$ in the following form:
\begin{align}
S^A & = S^A_{conf} + {S^A_{int}} \label{eq:SA}\\
S^A_{conf} & = -k_B\sum_{j=1}^{N_{conf}^A}p_jlnp_j \\
{S^A_{int}} & = \sum_{j=1}^{N_{conf}^A}p_jS_j^A 
\end{align}
\begin{align}
S^B & = S^B_{conf} + {S^B_{int}} \label{eq:SB}\\
S^B_{conf} & = -k_B\sum_{k=1}^{N_{conf}^B}q_klnq_k \\
{S^B_{int}} & = \sum_{k=1}^{N_{conf}^B}q_kS_k^A
\end{align}
With a simple algebraic manipulation shown below:
\begin{align}
S_{conf}^A & = -k_B\sum_{j=1}^{N_A}p_j\left( lnp_j - lnN_{conf}^A + lnN_{conf}^A\right) \nonumber \\
                     & = k_BlnN_{conf}^A - k_B\sum_{j=1}^{N_{conf}^A}p_jlnN_{conf}^Ap_j \label{SconfA}
\end{align}
Conformational entropy of macrostate $A$ ($S^A_{conf}$) is divided into two terms. The first term is the Boltzmann entropy (or ideal gas entropy, denoted as $S_{Boltzmann}^A$) based on the number of conformers. The second term represents deviation from the Boltzmann entropy (denoted as $\delta S_{conf}^A$). It is the product of the Boltzmann constant and the Kullback-Leibler divergence\cite{kullback1951} between the actual probability distribution of conformers in macrostate $A$ (${\bf{p}} = (p_1, p_2, \cdots, p_{N_{conf}^A})$) and the uniform distribution ($unif\{1,N_{conf}^A\}$). $S_{conf}^A$ may be rewritten as:
\begin{align}
S_{conf}^A & = S_{Boltzmann}^A + \delta S_{conf}^A \label{eq:SBconfA}\\
\delta S_{conf}^A & = -k_BD_{KL}({\bf{p}}||unif\{1,N_{conf}^A\}) \label{eq:dscA}
\end{align}
Similarly, denote probability distribution of conformers in macrostate $B$ as ${\bf{q}} = (q_1, q_2, \cdots, q_{N_{conf}^B})$ and the corresponding uniform distribution as $unif\{1,N_{conf}^B\}$, we have:
\begin{align}
S_{conf}^B & = S_{Boltzmann}^B + \delta S_{conf}^B \label{eq:SBconfB} \\
\delta S_{conf}^B & = -k_BD_{KL}({\bf{q}}||unif\{1,N_{conf}^B\}) \label{eq:dscB}
\end{align}
For ECISWD, if we denote the corresponding ISWD with a continuous probability density $\bf{R}$, then $\bf{p} \approx R $ and $ \bf{q} \approx R$. Denote the continuous uniform distribution as $\bf{unif}$, we have: 
\begin{align}
\delta S_{conf}^A \approx  & -k_BD_{KL}(\bf{R}||\bf{unif}) \\
\delta S_{conf}^B \approx & -k_BD_{KL}(\bf{R}||\bf{unif}) \\
\delta\Delta S_{conf}^{AB} & = \delta S^B_{conf} - \delta S^A_{conf}  \approx 0 \label{eq:dDS}\\
\Delta S_{conf}^{AB} & \approx k_Bln\frac{N^B_{conf}}{N^A_{conf}} \label{eq:sconf} 
\end{align}
Note that $\Delta S_{conf}^{AB}$ (equation \ref{eq:sconf}) is equivalent to $\Delta F^{AB}$ (equation \ref{eq:conf}) except a mere difference of a negative temperature factor. $\delta\Delta S_{conf}^{AB}$ reflect the difference between two KL divergences, which correspond to distances between the statistical weight distribution of conformers in macrostate $A(B)$ and the uniform distribution. The advantage of utilizing ECISWD for defining conformational entropy is the generality by concealing system specific molecular structural information in specific definition of conformers. Additionally, when difference of conformational entropy is taken between two arbitrary macrostates, deviation of the unknown ISWD from the uniform distribution is cancelled and we need only to deal with the number of conformers. Based on the same logic as in the case of free energy analysis,  with increasingly larger subsets of a sufficiently well-converged MD trajectory set, we expect to observe systematic decrease of $\delta\Delta S_{conf}$ calculated for arbitrarily defined macrostate pairs as long as ECISWD are basic states of conformational entropy. Conversely, we expect to observe systematic increase of $\delta\Delta S_{conf}$ when explicit conformers with widely variant statistical weight distributions are basic states of conformational entropy. To this end, we took the same trajectory sets, definition of conformers and macrostates as in the analysis of $\delta\Delta F$, and calculated corresponding $\delta\Delta S_{conf} = \delta S_{conf}^B - \delta S_{conf}^A$ based on equations (\ref{eq:dscA}) and (\ref{eq:dscB}) for each macrostate pair.  Both distributions of $\delta\Delta S_{conf}$ and corresponding CPD of its absolute value were shown in Fig. \ref{fig:SCPD}. As expected, and consistent with free energy analysis as shown in Fig. \ref{fig:CPD}, trend of $\delta\Delta S_{conf}$ based on conformers in set CONF1 (Fig. \ref{fig:SCPD}a,e) matches our expectation for that of ECISWD, while trends of $\delta\Delta S_{conf}$ based on conformers in sets CONF2 through CONF4 (Fig. \ref{fig:SCPD}bcd, fgh) match our expectation for that of conformers with variant statistical weight distribution, with coarser conformers and larger trajectory sets correspond to wider distributions of $\delta\Delta S_{conf}$.

\section*{Entropy enthalpy compensation}
In canonical ensemble, we have:
\begin{align}
\Delta F^{AB}& = \Delta U^{AB} - T\Delta S^{AB}  = \Delta U^{AB} - T(S^B - S^A) \label{eq:FEtot}
\end{align}
with $\Delta U^{AB}$ being the change of potential energy between the two macrostates $A$ and $B$. Let $\Delta S^{AB}_{int} = S^B_{int} - S^A_{int}$, and substitute equations \ref{eq:SA}, \ref{eq:SB}, \ref{eq:SBconfA}, \ref{eq:SBconfB} and \ref{eq:dDS} into equation (\ref{eq:FEtot}), we have:
\begin{align}
\Delta U^{AB} & \approx T\Delta S_{int}^{AB}\label{eq:eec}
\end{align}
While the derivation is carried out in canonical ensemble, it should be applicable for many isobaric-isothermal processes (e.g. many biomolecular systems under physiological conditions or routine experimental conditions) where change of the $PV$ term is negligible. Note that equation \ref{eq:eec} is the intriguing entropy-enthalpy compensation (EEC) phenomena (when the $PV$ term is negligible), which had long been an enigma\cite{Lumry1970,Imai1976,Grunwald1995,Gallicchio1998,Liu2001}, and has attracted a revival of interest due to its critical relevance in protein-ligand interactions\cite{Ford2005,Krishnamurthy2006,Krishnamurthy2006a,Starikov2007,Ward2010,Liu2011,Olsson2011,Ferrante2012,Starikov2012a,Chodera2013,Breiten2013,Tidemand2014,Ahmad2015}. Careful statistical analysis confirm that EEC does exist to various extent in many protein-ligand interaction systems after experimental errors are effectively removed\cite{Olsson2011}. 
For a given molecular system, once we have constructed a set of ECISWD, equations (\ref{eq:conf}) and (\ref{eq:eec}) state that change of molecular interactions does not necessarily cause change of free energy, which depends on relative number of thermally accessible ECISWD in end macrostates, and local effects from change of molecular interactions will be cancelled almost completely by corresponding change of average intra-conformer entropy. Note that correlation of neither signs nor magnitudes between $\Delta S_{conf}^{AB}$ and $\Delta S_{int}^{AB}$ is implied. Therefore, depending upon signs and magnitudes of $\Delta S_{conf}^{AB}$ and $\Delta S^{AB}_{int}$ (we neglect the $PV$ term here), this theory is compatible with molecular processes driven by enthalpy, entropy or both and various extent of observed EEC. When $\Delta S_{conf}^{AB} \approx 0$, perfect EEC would be observed; when $\Delta S_{conf}^{AB} > 0$ and $\Delta U > 0$ (or $\Delta S_{int}^{AB} > 0$), a seemingly entropy driven (and a reverse entropy limited) process would be observed; when $\Delta S_{conf}^{AB} > 0$ and $\Delta U < 0$ (or $\Delta S_{int}^{AB} < 0$), depending upon the sign of $\Delta S^{AB} = \Delta S^{AB}_{conf} + \Delta S^{AB}_{int}$ , a seemingly enthalpy or entropy-enthalpy jointly driven (and a reverse enthalpy or entropy-enthalpy jointly limited) process would be observed. The fundamental new perspective provided by equations (\ref{eq:conf}, \ref{eq:sconf} and \ref{eq:eec}) is that EEC is directly related to local redistribution of microstates in configurational space, while change of free energy and conformational entropy reflect the collective thermal accessibility of relevant macrostates. 
System complexity is essential for construction of ECISWD as demonstrated by our initial discussions on the double well model. Consistently, robustness of approximations in equations (\ref{eq:conf}) and (\ref{eq:sconf}) corresponds to the near-perfect cancellation of change of intra-conformer entropy and change of enthalpy as reflected by equations (\ref{eq:eec}). Without sufficient number of complex and heterogeneous microstates within each conformer, it is hard to imagine how such EEC occur. Along the same lines, a simple Morse potential type of protein-ligand interaction model was not found to allow significant EEC\cite{Chodera2013}. Based on the widespread observation of strong EEC effect in many molecular systems, it was suggested\cite{Chodera2013} that any attempt to calculate the change of free energy as a sum of its enthalpic and entropic contributions is likely to be unreliable. The proposed conformer counting strategy (equation \ref{eq:conf}) implicitly utilizes EEC by completely avoiding direct calculation of $\Delta U$ and $\Delta S_{int}$, which is expensive and error prone.  
\section*{Conclusions}
In summary, we presented the idea that snapshots in a converged MD trajectory set map directly to implicit thermally accessible conformers with ISWD. Based on the thought that infinite number of ways exist for defining implicit conformers with ISWD for a given molecular system, we hypothesized that any sufficiently fine set of conformers should have sufficiently good approximate ISWD. This hypothesis, while being disproved by a double well potential, tested successfully on extensive MD trajectories of lipid POPC. We think that competition of many DOFs, each allowed to vary significantly in both potential energy and spatial position within a conformer, constitutes the foundation for the observed validity of the hypothesis. Considering the moderate complexity of lipid POPC, it is likely that the hypothesis holds for complex molecular systems in general. This is a useful demonstration of the idea that ``More is different''\cite{moreisdifferent}. Active research is undergoing in our group toward defining ECISWD for more biomolecular systems (e.g. protein-ligand, protein-protein interaction and protein-nucleic acid interactions systems with explicit or implicit solvation). Furthermore, when ECISWD are utilized as basic states for definition of conformational entropy, change of which between two macrostates was found to be equivalent with corresponding change of free energy except a mere difference of a negative temperature factor. Meanwhile, change of potential energy between two macrostates was found to cancel corresponding change of average intra-conformer entropy. This finding suggests that EEC is inherently a local phenomenon in configurational space, and is likely universal in complex molecular systems. While providing an alternative perspective to the long-standing enigmatic EEC, this result is consistent with different extent of EEC observed for both enthalpy driven and entropy driven molecular processes in conventional sense where change of enthalpy is compared with change of total entropy. Counting thermally accessible ECISWD (equation \ref{eq:conf}) is a natural extension of the population based free energy formula (equation \ref{eq:snap}), which is only useful posterior to a converged simulation. However, equation \ref{eq:conf} effectively utilizes EEC implicitly through separation of entropy into conformational entropy based on ECISWD and intra-conformer entropy, and renders direct utilization of SMC and importance sampling possible for rapid free energy difference estimation\cite{zhang2003,zhang2006}. In accordance with ``no free lunch theorem''\cite{nofreelunch}, this expected gain in efficiency pays the price of all dynamical and pathway information associated with converged trajectories. 

\section*{Methods}
\begin{figure}[t] 
\centering 
\includegraphics[width=5.0in]{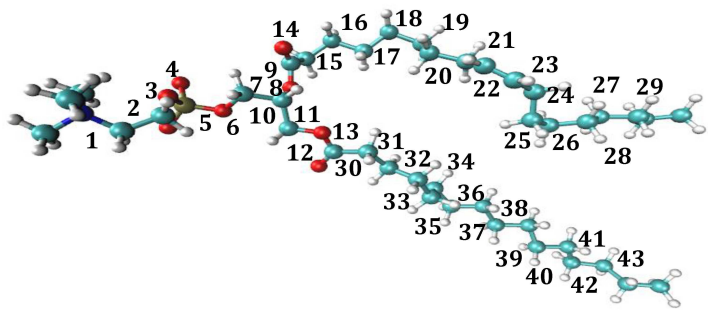}
\caption{Ball and stick representation of POPC and definition of conformer sets. Oxygen: red, hydrogen: white, carbon: cyan, phosphate: blue. The 43 all-heavy-atom torsions (see Table \ref{t:43torsion} for detailed lists of comprising atoms) utilized to define conformers are labeled with numbers on their central bonds. Set CONF1 is defined with all 43 torsions; set CONF2 is defined by 28 torsions, which are \{2,3,5,6,8,9,11,12,14,15,17,18,20,21,23,24,26,27,29,30,32,33,35,36,38,39,41,42\}; set CONF3 is defined by 22 odd numbered torsions and set CONF4 is defined by 15 torsions that are excluded in the definition of CONF1.}
\label{fig:POPC}
\end{figure}

To define conformers, we first take a given set of torsional DOFs (Fig. \ref{fig:POPC}), with each being divided into three equally sized torsional states with boundaries at $0^{\circ}(360^{\circ})$, $120^{\circ}$ and $240^{\circ}$, and subsequently utilize their unique combinations as conformers. Two structural states (i.e. snapshots) of a POPC molecule belong to the same conformer if and only if they share the same torsional state for each selected torsional DOF. 
Apparently, infinite number of ways exist to define set of conformers with similar fineness and uniformity.

To prepare macrostates, all snapshots in a given trajectory set were projected onto a selected backbone dihedral that was partitioned into 20 $18^{\circ}$-windows, snapshots fall within each of which constitute an observed macrostate. Such projections were performed for each of 43 dihedrals (Fig. \ref{fig:POPC}) and we have collectively 860 macrostates for each given combination of trajectory set and conformer definition. Apparently, macrostates based on the same dihedral angle do not overlap, while those based on different dihedral angles may overlap to different extent. To assign each snapshots to its belonging conformer and calculate $N_{conf}$ for each constructed macrostates, torsional states for the selected torsional DOFs were encoded into bit vectors and the radix sort algorithm\cite{IntroAlgorithm} was utilized.

Trajectory sets TSA1, TSA2 and TSA3 are constructed from snapshots of POPC collected in simulation condition A in the supplementary table 2 of the GPCR simulation study\cite{Dror2013}. There were totally 34143653 snapshots, which collectively amount to $\sim 6.15ms$ ($6.14585754ms$). Five subsets, with collective length (CL) being $\sim1.58ms$, $\sim1.32ms$, $\sim1.32ms$, $\sim1.32ms$ and $\sim0.66ms$ respectively, were available for this simulation condition. We take the first six trajectories out of the total 66 trajectories of the first subset as TSA1, which has a CL of $142.56\mu s$. The first subset ($\sim1.58ms$) was taken as TSA2, and the union of all subsets was taken as TSA3 ($\sim6.15ms$).  

\begin{acknowledgement}
This research was supported by National Natural Science Foundation of China under grant number 31270758, and by the Research fund for the doctoral program of higher education under grant number 20120061110019. Computational resources were partially supported by High Performance Computing Center of Jilin University, China. We thank DE Shaw Research for providing trajectory sets. We thank Zhonghan Hu for insightful discussions.
\end{acknowledgement}


\bibliography{../FreeEnergy3,../Total}


\newpage
\begin{table}
\centering
\begin{tabular}{|c|c|c|c|c|c|c|c|c|c|}
\hline
Index&atom1&atom2&atom3&atom4&Index&atom1&atom2&atom3&atom4\\
\hline
1&C12&N&C11&C15&2&N&C11&C15&O1\\
\hline
3&C11&C15&O1&P1&4&C15&O1&P1&O2\\
\hline
5&O1&P1&O2&C1&6&P1&O2&C1&C2\\
\hline
7&O2&C1&C2&O21&8&C1&C2&O21&C21\\
\hline
9&C2&O21&C21&C22&10&O2&C1&C2&C3\\
\hline
11&C1&C2&C3&O31&12&C2&C3&O31&C31\\
\hline
13&C3&O31&C31&C32&14&O21&C21&C22&C23\\
\hline
15&C21&C22&C23&C24&16&C22&C23&C24&C25\\
\hline
17&C23&C24&C25&C26&18&C24&C25&C26&C27\\
\hline
19&C25&C26&C27&C28&20&C26&C27&C28&C29\\
\hline
21&C27&C28&C29&C210&22&C28&C29&C210&C211\\
\hline
23&C29&C210&C211&C212&24&C210&C211&C212&C213\\
\hline
25&C211&C212&C213&C214&26&C212&C213&C214&C215\\
\hline
27&C213&C214&C215&C216&28&C214&C215&C216&C217\\
\hline
29&C215&C216&C217&C218&30&O31&C31&C32&C33\\
\hline
31&C31&C32&C33&C34&32&C32&C33&C34&C35\\
\hline
33&C33&C34&C35&C36&34&C34&C35&C36&C37\\
\hline
35&C35&C36&C37&C38&36&C36&C37&C38&C39\\
\hline
37&C37&C38&C39&C310&38&C38&C39&C310&C311\\
\hline
39&C39&C310&C311&C312&40&C310&C311&C312&C313\\
\hline
41&C311&C312&C313&C314&42&C312&C313&C314&C315\\
\hline
43&C313&C314&C315&C316&&&&&\\
\hline
\end{tabular}
\caption{Detailed list of comprising atoms of the 43 torsions utilized in defining conformers and macrostates for POPC.}
\label{t:43torsion}
\end{table}

\newpage

\newpage

\newpage

\newpage
\captionsetup[subfigure]{labelformat=empty}

\newpage

\newpage
\begin{figure}[] 
\centering 
\subfloat[]{\includegraphics[width=1.6in]{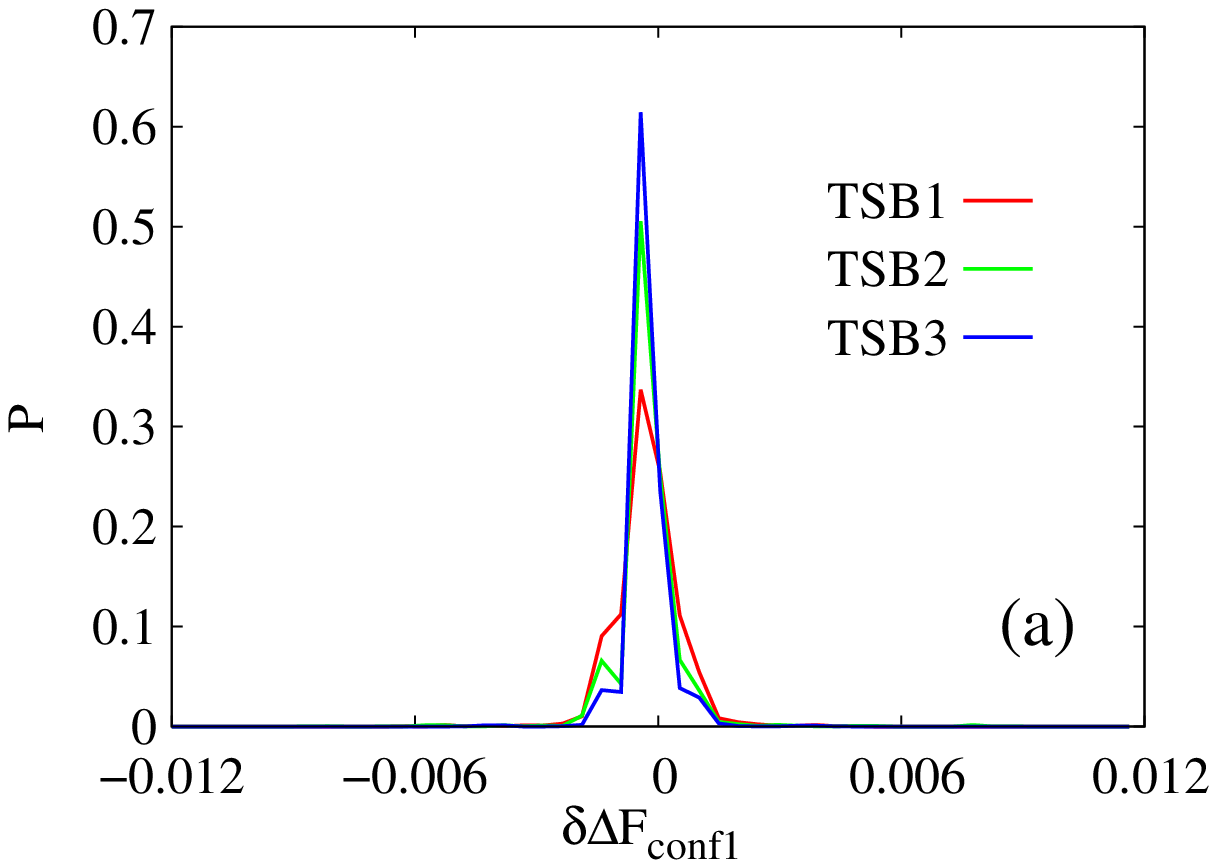}}
\subfloat[]{\includegraphics[width=1.6in]{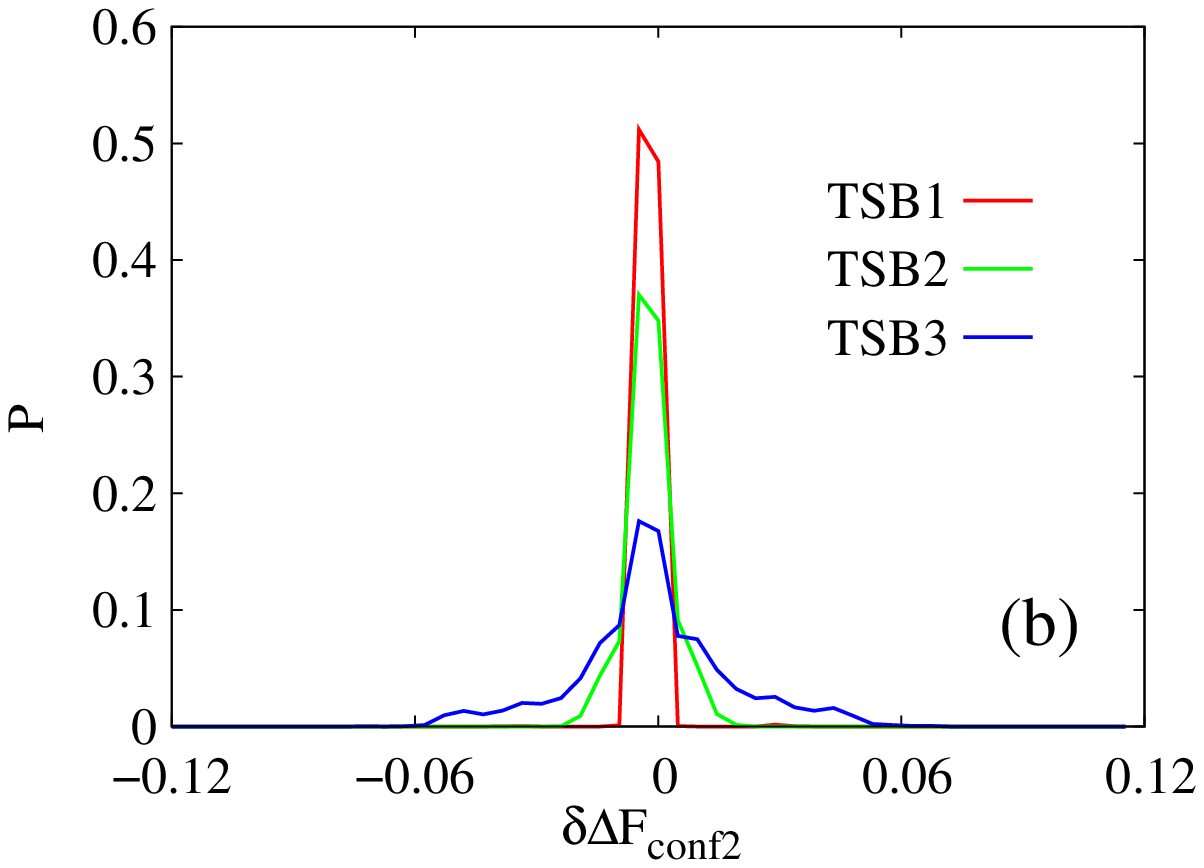}}
\subfloat[]{\includegraphics[width=1.6in]{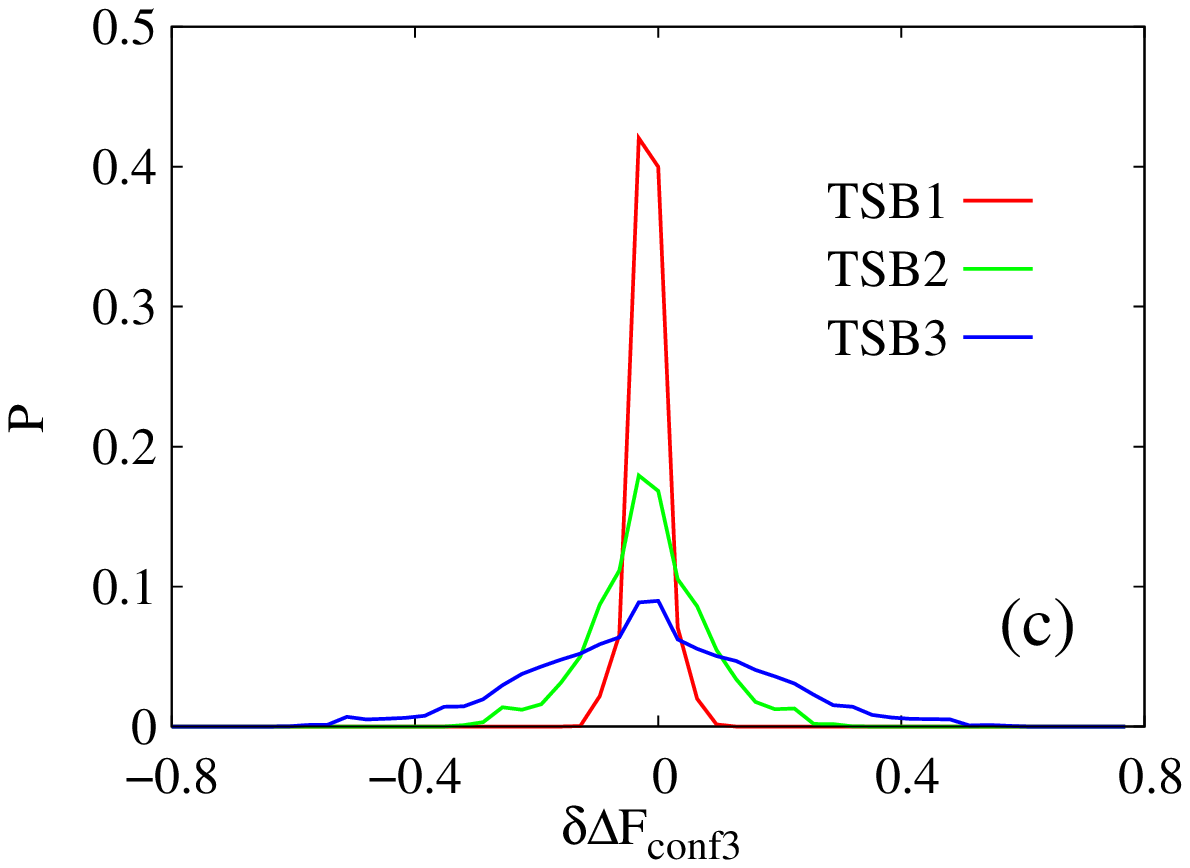}}
\subfloat[]{\includegraphics[width=1.6in]{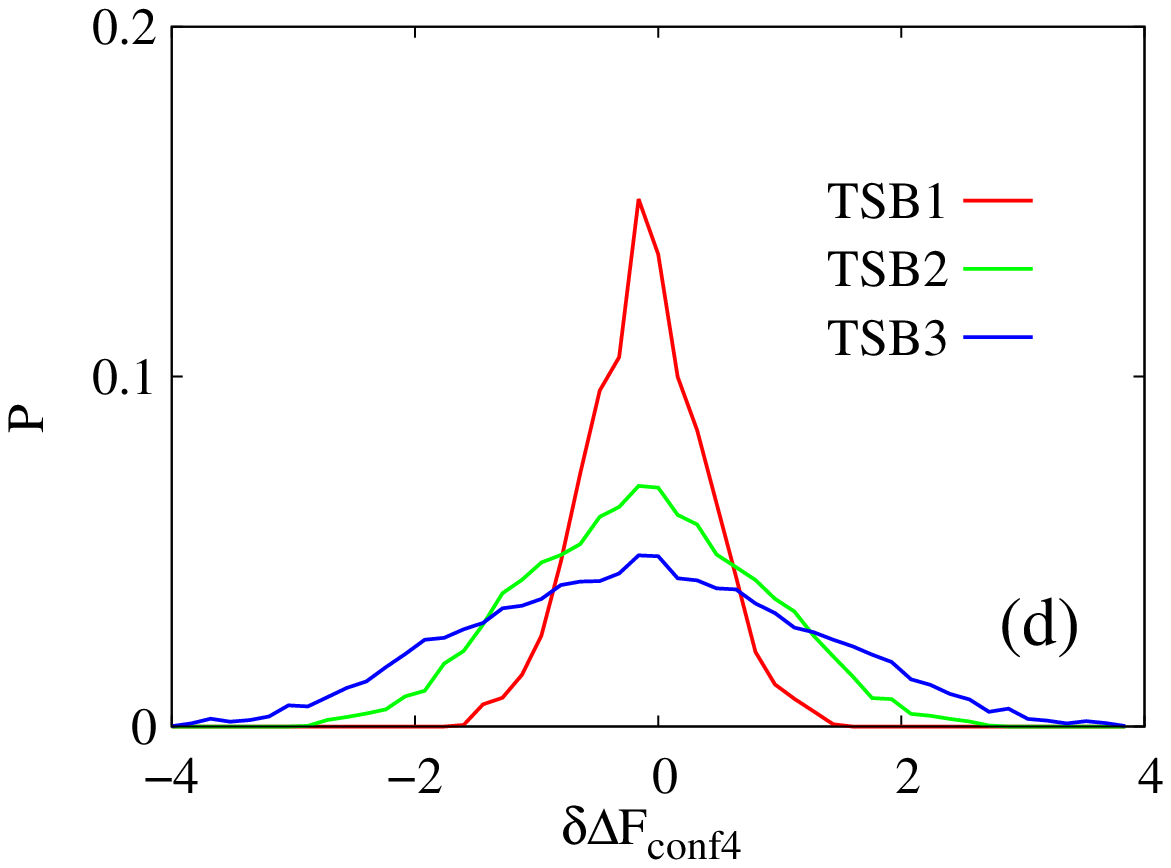}}\\
\subfloat[]{\includegraphics[width=1.6in]{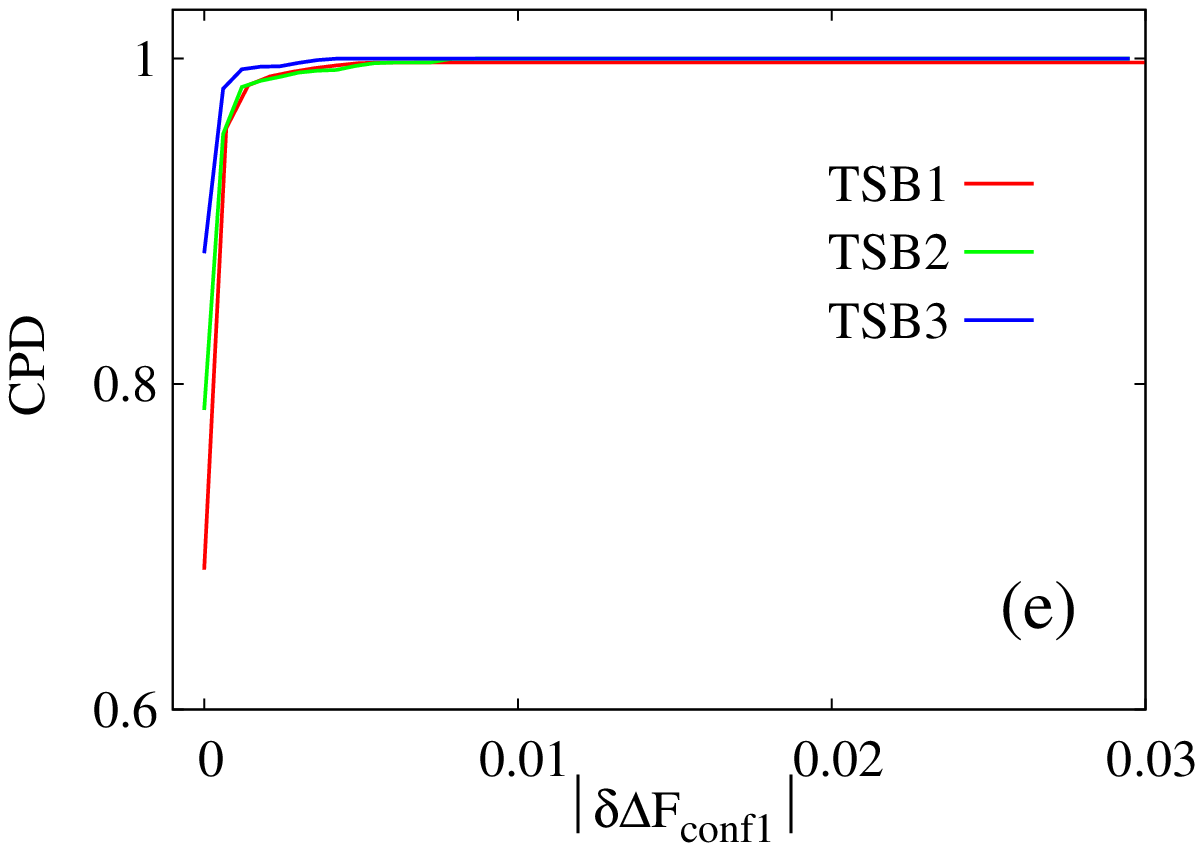}}
\subfloat[]{\includegraphics[width=1.6in]{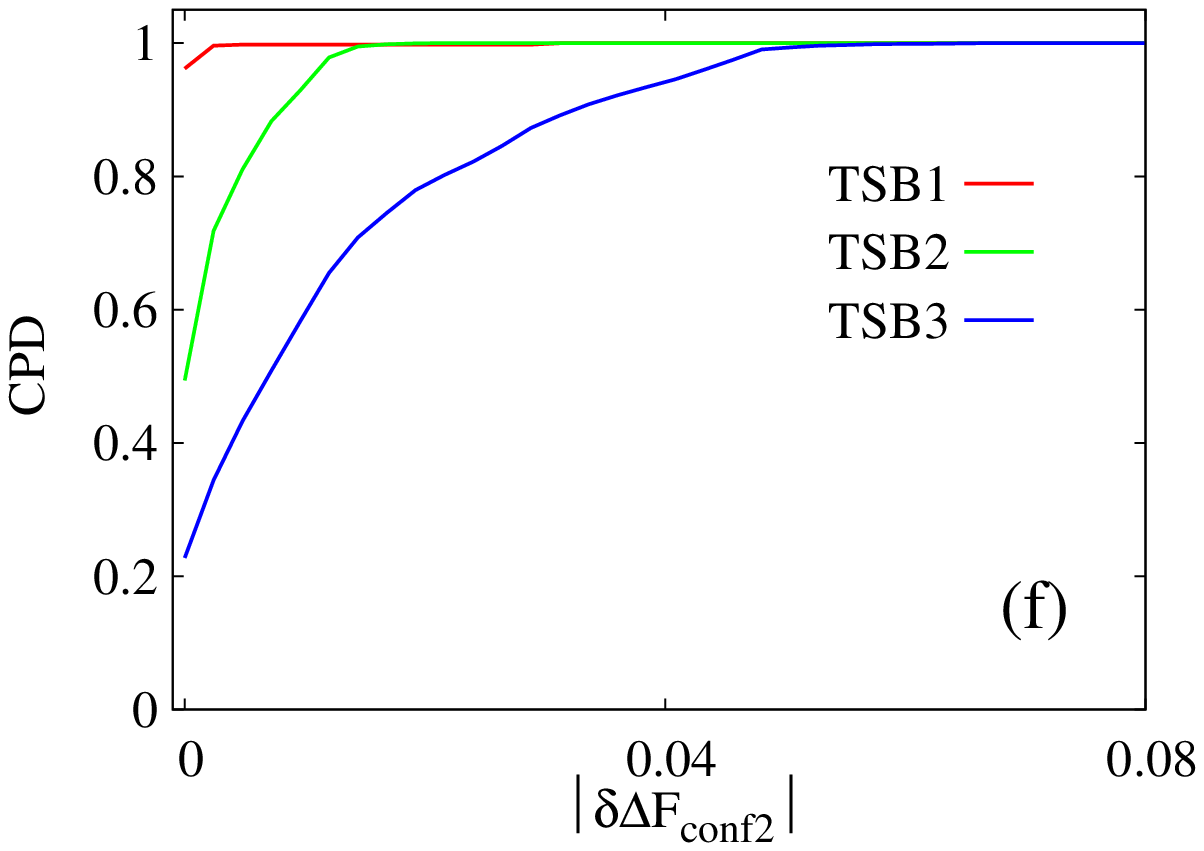}}
\subfloat[]{\includegraphics[width=1.6in]{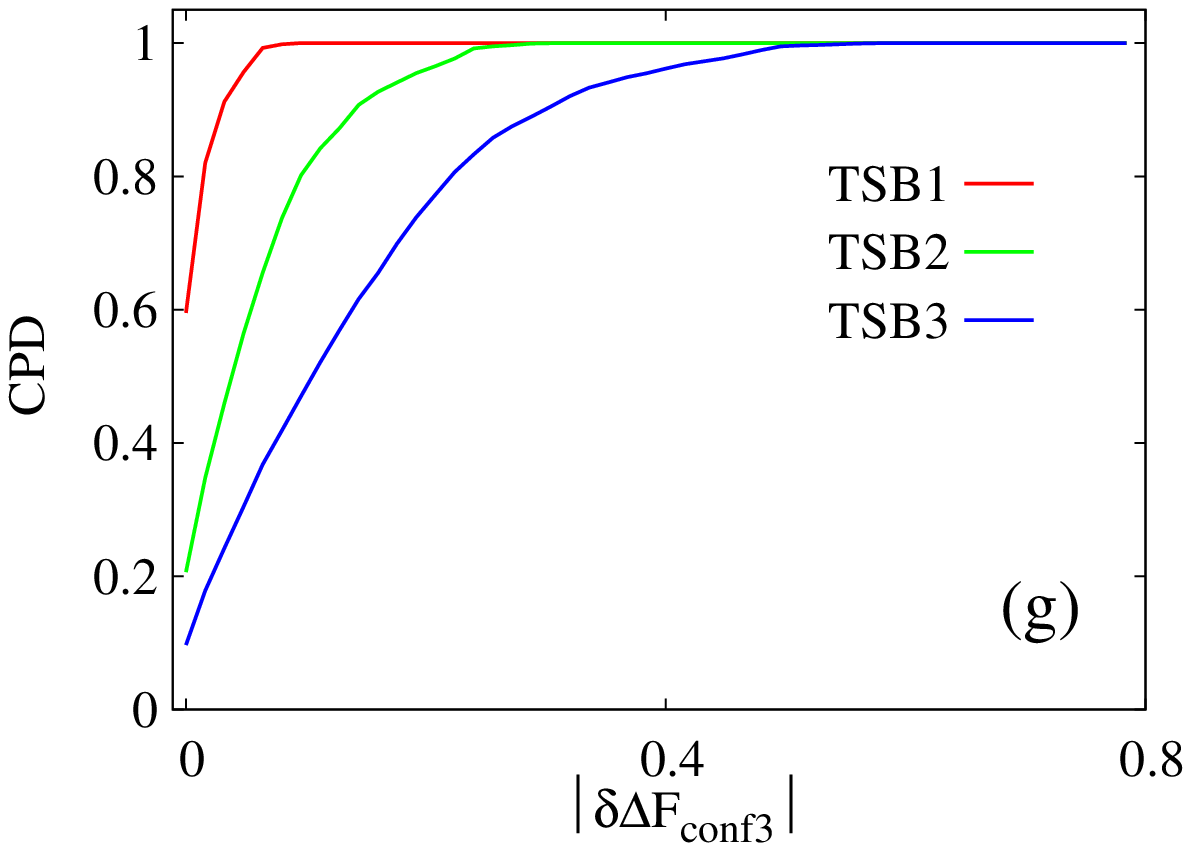}}
\subfloat[]{\includegraphics[width=1.6in]{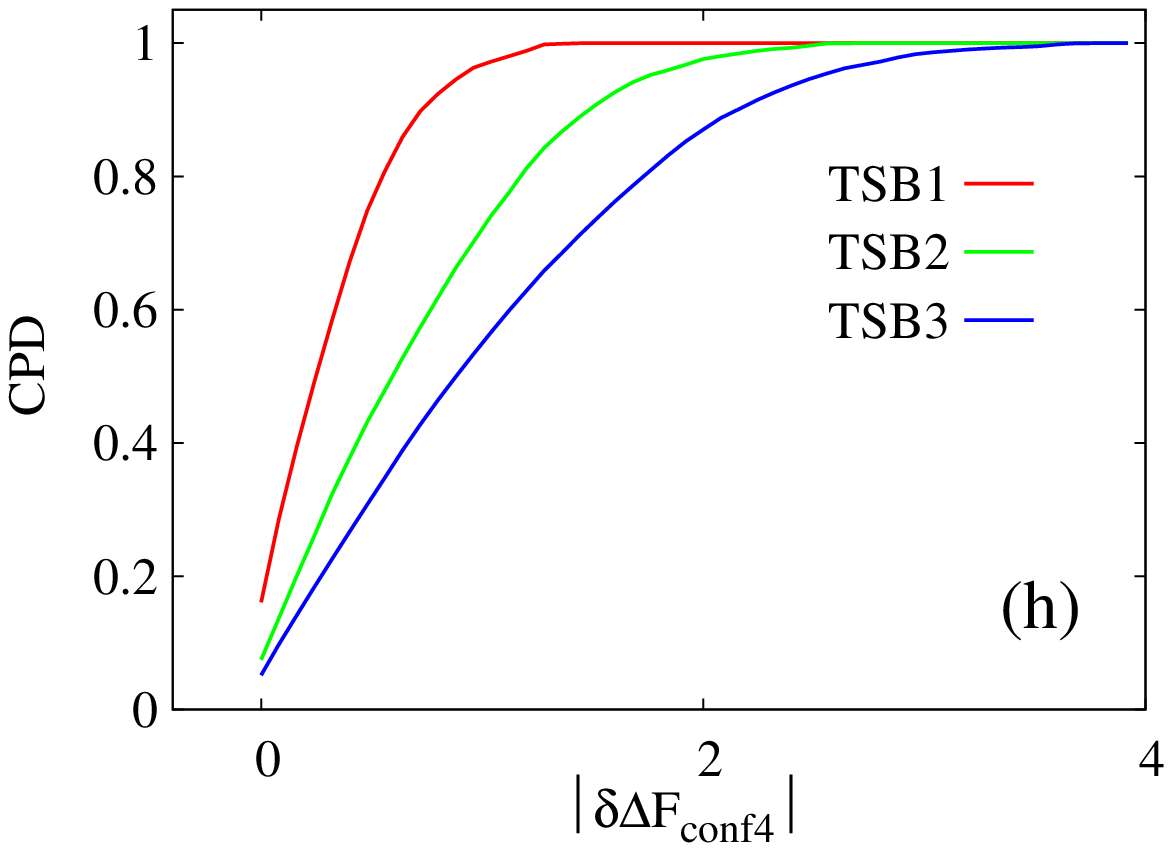}}\\
\caption{Distributions of $\delta\Delta F$ (a - d) and CPD  of its absolute values (e - h) for POPC with conformer sets CONF1 through CONF4 on trajectory sets TSB1 through TSB3. These trajectory sets are constructed from snapshots of POPC collected in simulation condition B in the supplementary table 2\cite{Dror2013}. There were 36724760 snapshots, which collectively amount to a CTS of $\sim 6.61ms$ ($6.6104568ms$). Five subsets, each including 56 trajectories with CTS being $\sim1.32ms$, were available for this simulation condition. After trajectories of the first subset were sorted according to file name, the first six trajectories were taken as TSB1 ($\sim 200\mu s$). The first subset is taken as TSB2 ($\sim 1.32ms$), and the union of all subsets was taken as TSB3 ($\sim6.61ms$).  Different trajectory sets are represented by different line colors. The unit of the horizontal axis is in $k_BT$.} 
\label{fig:TSB}
\end{figure}

\newpage

\newpage

\end{document}